\documentclass[
  aps,
  prl,
  reprint,
  superscriptaddress,
  nofootinbib
]{revtex4-2}

\usepackage[T1]{fontenc}
\usepackage[utf8]{inputenc}
\usepackage{amsmath,amssymb,bm}
\usepackage{graphicx}
\usepackage{xcolor}
\usepackage{booktabs}
\usepackage{multirow}
\usepackage{makecell}
\usepackage{hyperref}

\usepackage{adjustbox}

\begin{document}


\title{Milky Way Dynamics Favor Dark Matter over Modified Gravity Models}

\author{Zheng-long Wang}
\affiliation{Key Laboratory of Dark Matter and Space Astronomy, Purple Mountain Observatory, Chinese Academy of Sciences, Nanjing 210033, People's Republic of China}
\affiliation{School of Astronomy and Space Science, University of Science and Technology of China, Hefei, Anhui 230026, People's Republic of China}

\author{Yue-Lin Sming Tsai}
\email{smingtsai@pmo.ac.cn}
\affiliation{Key Laboratory of Dark Matter and Space Astronomy, Purple Mountain Observatory, Chinese Academy of Sciences, Nanjing 210033, People's Republic of China}
\affiliation{School of Astronomy and Space Science, University of Science and Technology of China, Hefei, Anhui 230026, People's Republic of China}

\author{Lan Zhang}
\email{zhanglan@nao.cas.cn}
\affiliation{National Astronomical Observatories, Chinese Academy of Sciences, Beijing 100012, People's Republic of China}

\author{Yin Wu}
\affiliation{National Astronomical Observatories, Chinese Academy of Sciences, Beijing 100012, People's Republic of China}
\affiliation{School of Astronomy and Space Science, University of Chinese Academy of Sciences, Beijing 100049, People's Republic of China}

\author{Haining Li}
\affiliation{National Astronomical Observatories, Chinese Academy of Sciences, Beijing 100012, People's Republic of China}

\author{Xiang-Xiang Xue}
\affiliation{National Astronomical Observatories, Chinese Academy of Sciences, Beijing 100012, People's Republic of China}

\author{Hongsheng Zhao}
\email{hz4@st-andrews.ac.uk}
\affiliation{Scottish Universities Physics Alliance, University of Saint Andrews, North Haugh, Saint Andrews, Fife, KY16 9SS, UK}
\affiliation{Department of Astronomy, School of Physical Sciences, University of Science and Technology of China, Hefei, Anhui 230026, China}
\affiliation{Institute for Advanced Study in Physics, Zhejiang University, Hangzhou 310058, China}

\author{Yi-Zhong Fan}
\email{yzfan@pmo.ac.cn}
\affiliation{Key Laboratory of Dark Matter and Space Astronomy, Purple Mountain Observatory, Chinese Academy of Sciences, Nanjing 210033, People's Republic of China}
\affiliation{School of Astronomy and Space Science, University of Science and Technology of China, Hefei, Anhui 230026, People's Republic of China}

\date{\today}

\begin{abstract}
Modified gravity theories such as Modified Newtonian Dynamics (MOND) and Scalar-Tensor-Vector Gravity (STVG) have been proposed as alternatives to dark matter, but decisive tests have been hindered by degeneracies between baryonic structure and gravitational laws.
Here we break this degeneracy using independent, high-precision constraints: the Milky Way radial rotation curve, vertical phase-space spirals from Gaia, and a broken-exponential stellar disk.
A joint reconstruction of the radial and vertical gravitational fields reveals a structural inconsistency in modified gravity---no model can simultaneously reproduce both observations.
Our results strongly disfavor MOND at $>13\sigma$ and STVG at $>4\sigma$.
In contrast, dark matter halo models naturally explain the observations, providing a self-consistent test of gravity on galactic scales.
\end{abstract}

\maketitle


\textit{Introduction.---}
The "small-scale crisis"~\cite{Bullock_2017} of the standard cold dark matter (CDM) paradigm is potentially explained by gravity theories such as Modified Newtonian Dynamics (MOND)~\cite{Banik_2022}. Instead of non-baryonic matter, MOND introduces a fundamental acceleration scale $a_0$~\cite{Milgrom_1983,Famaey_2012}. Below this threshold, gravity is non-linearly enhanced to mimic a phantom dark halo without cuspy density and dynamical friction.
On galactic scales, MOND demonstrates remarkable predictive power, having yielded a priori predictions for the Radial Acceleration Relation (RAR)~\cite{McGaugh_2016}, the cored halos in dwarf galaxies~\cite{McGaugh_2020}, and the long-lived bars in high brightness galaxies~\cite{Roshan_2021}. Other modified gravity theories, such as Scalar--Tensor--Vector Gravity (STVG), have also been proposed to account for similar galactic-scale phenomena~\cite{Moffat_2013,Green_2019}. 
Recent studies have examined MOND-related scenarios in Gaia wide binaries~\cite{Chae:2023prf,chae2026}, the KBC void and the Hubble tension~\cite{Haslbauer_2020}, and the cosmic microwave background~\cite{Angus_2009}. Related work has explored MOND in dwarf-galaxy systems~\cite{Bilek_2025}, tested MOND-inspired cosmologies with large-scale simulations~\cite{Russell_2026}, and reported tensions for Local Group satellite systems within $\Lambda$CDM simulations~\cite{Haslbauer_2024}. However, such claims remain debated~\cite{Banik:2023pbo}. Beyond galaxy scales, the MOND extension is non-unique, whereas CDM is well established~\cite{Alam_2017,Aghanim_2020}.
The decisive test for MOND and STVG is still in the galactic details, especially with the new kinds of data from Gaia.

Previously, testing dark matter and gravity models against Milky Way (MW) observational data has been hampered by several challenges. 
In the absence of direct measurements of the inner disk stellar distribution, 
one has to extrapolate profiles from the outer disk~\cite{Juri__2008}, 
thereby sustaining a degeneracy between baryonic and gravitational models (dark matter halos or modified gravity). 
This degeneracy can be broken by introducing independent vertical kinematic constraints. 
However, traditional methods employing the vertical Jeans equation \cite{Nipoti_2007, Lisanti_2019, Davari_2020,Zhu_2022,Lopez_2025} assume steady-state equilibrium, 
a premise now invalidated due to ongoing vertical perturbations. These perturbations, most notably manifested as phase-space ``snail'' structures \cite{Antoja_2018}, 
introduce systematic uncertainties into equilibrium models, 
ultimately rendering their conclusions unreliable. 

Our current investigation benefits from two breakthroughs. 
First, the phase-space ``snail'' structure allows us to explicitly account for non-equilibrium dynamics, achieving a high-precision, model-independent measurement of the vertical potential that completely bypasses steady state assumptions~\cite{Guo_2024}. 
Second, and crucially, an unprecedentedly accurate measurement of the MW baryonic distribution (particularly the stellar disk) is now achievable.
Analyses of Mono-Abundance Populations (MAPs) from spectroscopic surveys such as APOGEE~\cite{Bovy_2016, Mackereth_2017, Lian_2022, Cantat_2024, Imig_2025} and LAMOST~\cite{Yu_2021, Yu_2025} have unveiled that the MW stellar disk follows a complex broken-exponential profile, featuring a flat inner density that contrasts with traditional single-exponential models. This accurate density profiling yields a Galactic half-light radius of $5.75\pm 0.38$~kpc, consistent with local galaxies of similar mass~\cite{Lian_2024}. Meanwhile, the flat inner region significantly reduces the stellar mass in the inner MW, revealing that the gravitational acceleration provided by baryonic matter is much lower than previously estimated~\cite{Lian_2025}, thereby posing a  challenge to modified gravity theories. 
Thanks to these significant improvements, the dark matter and modified gravity models can be convincingly distinguished. 

\textit{Likelihoods and Gravitational Potential Reconstruction.---}
To break the degeneracy between baryonic mass and dark matter (or modified gravity) profiles, 
we construct a multi-dimensional dataset that jointly incorporates radial and vertical dynamical constraints. 
Using a cylindrical coordinate system $(R, \phi, z)$, we combine the rotation curve $v_c$ (a tracer of the radial gravitational field) and 
the vertical potential $\Phi_z$ (a probe of the vertical restoring force) in the MW.
By fitting these data with gravitational models containing both baryonic and non-baryonic components, 
we can assess which framework (CDM or modified gravity) more consistently explains MW dynamics.

The radial dynamics are constrained by a dataset of 120,309 Red Giant Branch (RGB) stars~\cite{Ou_2024}, 
which integrates \textsc{APOGEE DR17} spectroscopy~\cite{Majewski_2017} with multi-band photometry from Gaia DR3~\cite{Gaia_2023}, \textsc{2MASS}~\cite{2MASS}, and \textsc{WISE}~\cite{Wright_2010}.
The distances to these stars are determined via data-driven spectrophotometric parallaxes, achieving a 40\% improvement in precision over the original Gaia DR3 measurements~\cite{Ou_2024}.
We restrict our thin-disk sample to the anti-center sector ($150^{\circ} \le \phi \le 210^{\circ}$) across a radial range of $6 < R < 27.5$ kpc. Furthermore, we derive the rotation curve via the Jeans equation using a broken-exponential stellar number density profile~\cite{Lian_2022}. 
This approach accounts for disk structural complexities and corrects systematic biases inherent 
in previous single-exponential stellar number density profile~\cite{Eilers_2019, Zhou_2023, Ou_2024}, 
ensuring a more accurate dynamical reconstruction among the entire observed disk.

In the context of vertical dynamics, we introduce the vertical gravitational potential $\Phi_z$ as a physical constraint orthogonal to the radial rotation curve. Utilizing the Gaia DR3 radial velocity sample~\cite{Gaia_2023}, we reconstruct the local $\Phi_z$ from the $z-V_z$ phase-space snail structure (following the method of~\cite{Guo_2024}). By exploiting the vertical energy conservation relation
\begin{equation}
E_z = \Phi_z(z_{\mathrm{max}}) = \frac{V_{z,\mathrm{max}}^2}{2},
\label{eq:vertical_energy}
\end{equation}
this method directly maps the stellar phase-space geometric features---which are governed by non-equilibrium dynamics---into gravitational potential information. Unlike traditional vertical Jeans equation methods relying on the assumption of dynamical equilibrium~\cite{Buch_2019,Cheng_2024,Soding_2025}, this mapping strategy is model-independent and requires neither steady-state assumptions nor stellar density distributions. Consequently, we present independent measurements of the vertical gravitational field within the range of $7.8 < R_{\mathrm{pot}} < 11.0~\mathrm{kpc}$, where $R_{\mathrm{pot}}$ denotes the effective radius corrected for asymmetric drift.

For robustness, we explicitly evaluate potential systematic errors for both the rotation curve and the vertical potential. 
For the rotation curve, we account for biases owing to departures from the axisymmetric Jeans equation assumptions (specifically axisymmetry and dynamical equilibrium), tracer biases, and asymmetric drift. 
Thus, we adopt a systematic uncertainty that varies with radius $R$ from $5\%$ to $20\%$, i.e., the largest one reported in~\cite{Ou_2025}. 
For the vertical potential, simulation tests indicate a systematic uncertainty of at most $\sim 5\%$~\cite{Guo_2024}. 
In the final fitting process, the total error budget $\sigma_{\mathrm{total}}$ is constructed by adding the observational error $\sigma_{\mathrm{obs}}$ and the estimated systematic uncertainty $\sigma_{\mathrm{sys}}$ in quadrature, ensuring a conservative constraint on the model parameters.

The total baryonic mass model consists of three components, including a stellar disk, a bulge, and a gas disk.
The bulge and gas components are set to the most recent observational data~\cite{McMillan_2016,Sormani_2022}, given their relatively minor gravitational contribution within our study domain. For the stellar disk component, we adopt the comprehensive disk distribution derived from MAPs, stellar-evolution isochrones, and the initial mass function~\cite{Lian_2022}. This stellar distribution consists of thin and thick disks characterized by broken-exponential profiles. 
To reduce parameter degeneracies, we fix the geometric parameters of stellar disks to the latest tightly-constrained results \cite{Lian_2022}, leaving only the local solar‑neighbourhood densities of the thin and thick disks, namely $\rho_{\odot,\mathrm{thin}}$ and $\rho_{\odot,\mathrm{thick}}$ as free parameters in the fit.

Our main goal is to distinguish the dark matter and modified gravity models with these data.
For dark matter distribution, we adopt two special cases of the generalized Zhao model~\cite{Zhao_1996}, namely the canonical Navarro--Frenk--White (NFW) profile~\cite{Navarro_1997} and the more flexible Einasto profile~\cite{Einasto_1965}. 
For modified gravity, we examine both Quasi‑linear MOND (QUMOND) and STVG theory. 
For QUMOND, we examine three versions of the function $\nu(g_N/a_0)$, \textbf{Simple}, \textbf{Standard}, and \textbf{RAR}-inspired (see Supplementary Material). Here, $g_N$ denotes the Newtonian gravitational acceleration generated by the baryonic mass distribution.
Given the precise measurements from the Radial Acceleration Relation~\cite{McGaugh_2016}, we fix the characteristic acceleration scale $a_0$ to $1.2 \times 10^{-10} \mathrm{\, m \, s^{-2}}$.
In STVG, the dimensionless strength parameter $\alpha$ and the range parameter $\mu$, which are treated as free parameters in our analysis, determine the amplitude and characteristic scale of gravitational enhancement, respectively~\cite{Moffat_2006}.
To model the complex baryonic mass distribution and non-linear gravitational effects in QUMOND and STVG, we numerically solve the Poisson equation on a two-dimensional cylindrical grid within axisymmetric geometry.

We perform a global parameter-space scan for two dark matter profiles and four modified gravity models 
using a Markov Chain Monte Carlo (MCMC) algorithm. 
The likelihood function combines the rotation-curve data and the vertical gravitational-potential data. 
Equivalently, the total log-likelihood is written as
\begin{equation}
\ln \mathcal{L}_{\mathrm{tot}}
=
\ln \mathcal{L}_{\mathrm{rot}}
+
\ln \mathcal{L}_{\mathrm{vert}} .
\label{eq:Ltot}
\end{equation}
Details regarding the vertical potential modeling, along with the prior ranges and distributions, are provided in the Supplementary Material.
The model comparison is based on the full Bayesian evidence $\mathcal{Z}$. 
As independent checks, the reduced chi-square $\chi^{2}_{\mathrm{red}}$ and the Bayesian Information Criterion (BIC) are also calculated. 
In the Bayesian analysis, we apply Gaussian priors for the well-measured local stellar densities of the thin and thick disks~\cite{Lian_2025}. 
In contrast, for the frequentist metrics (the reduced $\chi^2$ and BIC), these same constraints are incorporated directly into the likelihood function.

\begin{figure}[t]
  \centering
  \includegraphics[width=\linewidth]{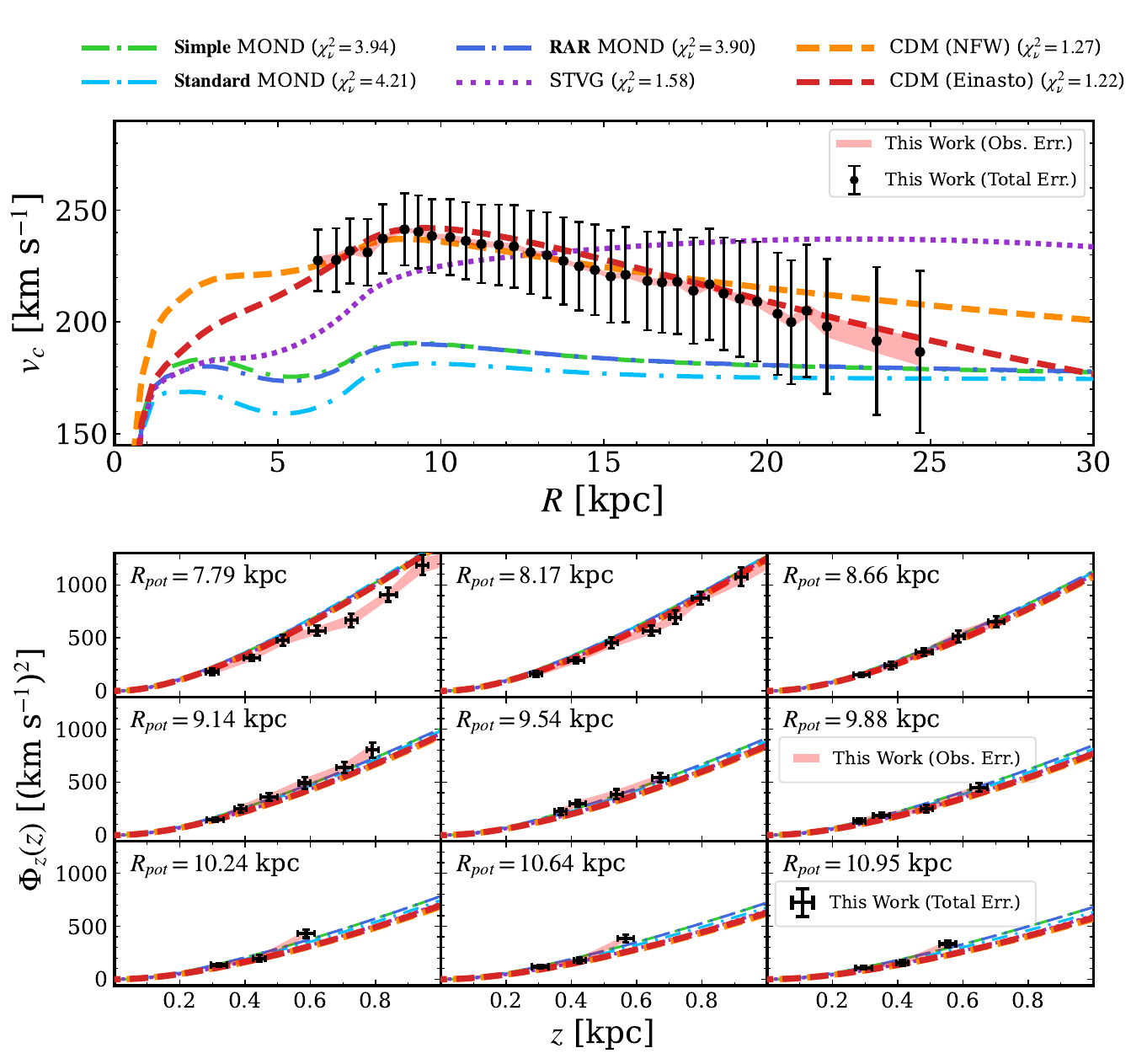}
  \caption{
  \textbf{The MW rotation curves (top panel) and vertical gravitational potential at selected radii (bottom panels) versus the model predictions.}
  The CDM models are represented by thick red dashed lines, whereas other colors denote distinct modified-gravity theories. While the CDM model successfully reproduces both dynamical constraints, modified gravity theories fail to match the observed rotation curve. Observational data points are derived from this work: black points with error bars represent the total error (observational plus systematic), whereas the red shaded regions indicate the observational error only. All model lines are just for the best-fit parameters.
  }
  \label{fig:main_radial_vertical}
\end{figure}

Details of the rotation curve construction, vertical potential reconstruction, gravitational modeling, likelihood construction, and statistical methodology are provided in the Supplemental Material.

\textit{Result.---}We present the rotation curves and vertical potentials in Fig.~\ref{fig:main_radial_vertical} (also see the fitted parameters organized in tables in the Supplementary Material). The CDM model reproduces both constraints within the broken-exponential disk framework. In contrast, modified gravity theories exhibit good agreement with the vertical potential but cannot accurately capture the intricate details of the rotation curve.

Our analysis prioritizes results incorporating conservative systematic errors to ensure robustness, while results based solely on observational errors are also included for comparison. The CDM model maintains a good fit even without systematic errors, whereas modified gravity models exhibit significantly poorer performance. Overall, when systematic errors are included, modified gravity remains strongly disfavored compared to CDM.
Specifically, QUMOND is excluded at $>13\sigma$ ($\Delta \text{BIC}>200$), and even the relatively better-performing STVG is disfavored at $4.2\sigma$ ($\Delta \text{BIC} = 24.0$). Based on Jeffreys' scale (where $\Delta \text{BIC} > 10$ implies decisive evidence against the model), even when simultaneously considering the maximum possible systematic and observational errors, both frameworks are effectively ruled out in the MW.

\begin{table}
	\centering
	\caption{\textbf{Model comparison between dark matter and modified gravity frameworks.}
		Columns give the reduced $\chi^2_{\nu}$ and Bayesian Information Criterion (BIC) for the best-fit models, followed by the relative log-evidence ($-\Delta \ln \mathcal{Z}$) and the statistical significance (in $\sigma$ units). Values without parentheses include both systematic and observational error, while values in parentheses show observational error only. The quantity $-\Delta \ln \mathcal{Z}$ measures the log-evidence difference relative to the most favored model, and determines the significance in the last column.}
	\label{tab:fit_results}
	
	\scriptsize
	\setlength{\tabcolsep}{3.5pt}
	\renewcommand{\arraystretch}{1.15}
	
	\begin{tabular}{lcccc}
		\hline
		Model & $\chi^2_{\nu}$ & $\Delta$BIC & $-\Delta\ln \mathcal{Z}$ & Sig. ($\sigma$) \\
		\hline
		\textbf{Simple} MOND    & 3.94 (40.54) & 202.5 (2995.7) & 92.3 (1464.6)  & 13.4 (54.0) \\
		\textbf{Standard} MOND  & 4.21 (64.01) & 223.5 (4825.9) & 108.6 (2400.7) & 14.5 (69.2) \\
		\textbf{RAR} MOND       & 3.90 (42.04) & 199.2 (3112.3) & 91.0 (1524.7)  & 13.3 (55.1) \\
		STVG                    & 1.58 (31.98) & 24.0 (2273.0)  & 10.5 (1136.0)  & 4.2 (47.6)  \\
		NFW                     & 1.27 (2.70)  & 0.7 (47.4)     & 0.6 (23.2)     & 0.9 (6.5)   \\
		Einasto                 & 1.22 (2.05)  & \textbf{0.0} (\textbf{0.0}) & \textbf{0.0} (\textbf{0.0}) & \textbf{0.0} (\textbf{0.0}) \\
		\hline
	\end{tabular}
\end{table}

Examining the posterior distributions including with all uncertainties shows that 
within the CDM framework, all parameters are well constrained except for the Einasto inner slope, 
which remains poorly determined due to rotation curve uncertainties.
Modified gravity theories, by contrast, exhibit severe observational tensions (see Supplementary Material). 
QUMOND predicts a local thin-disk density 
($\rho_{\odot,\thinspace\mathrm{thin}} \approx 2.5\times10^{-2}$ to $4.2\times10^{-2}~\mathrm{M_\odot\,pc^{-3}}$) 
that falls significantly below observational estimates~\cite{Lian_2025}, 
while STVG fitted parameters ($\alpha \approx 11.3$, $\mu \approx 0.068~\mathrm{kpc}^{-1}$) deviate remarkably from the universal galaxy-scale values ($\alpha \approx 8.89$, $\mu \approx 0.042~\mathrm{kpc}^{-1}$)~\cite{Moffat_2013}. 
Although this substantial departure enables STVG to accommodate the data with a modest exclusion significance ($\sim 4.2\sigma$), 
the non-universal parameter behavior reveals an inconsistency: 
the theory cannot simultaneously explain MW dynamics while maintaining predictive power across galactic systems.

\textit{Discussion.---}
Our analysis reveals a fundamental tension within modified gravity theories when simultaneously constrained by both radial and vertical dynamical data in the MW. 
Once the observed baryonic distribution is accounted for, the resulting gravitational potentials in these theories are dominated by the baryonic disk and fail to reproduce the observed kinematic data.

\begin{figure}
	\centering
	\includegraphics[width=\linewidth]{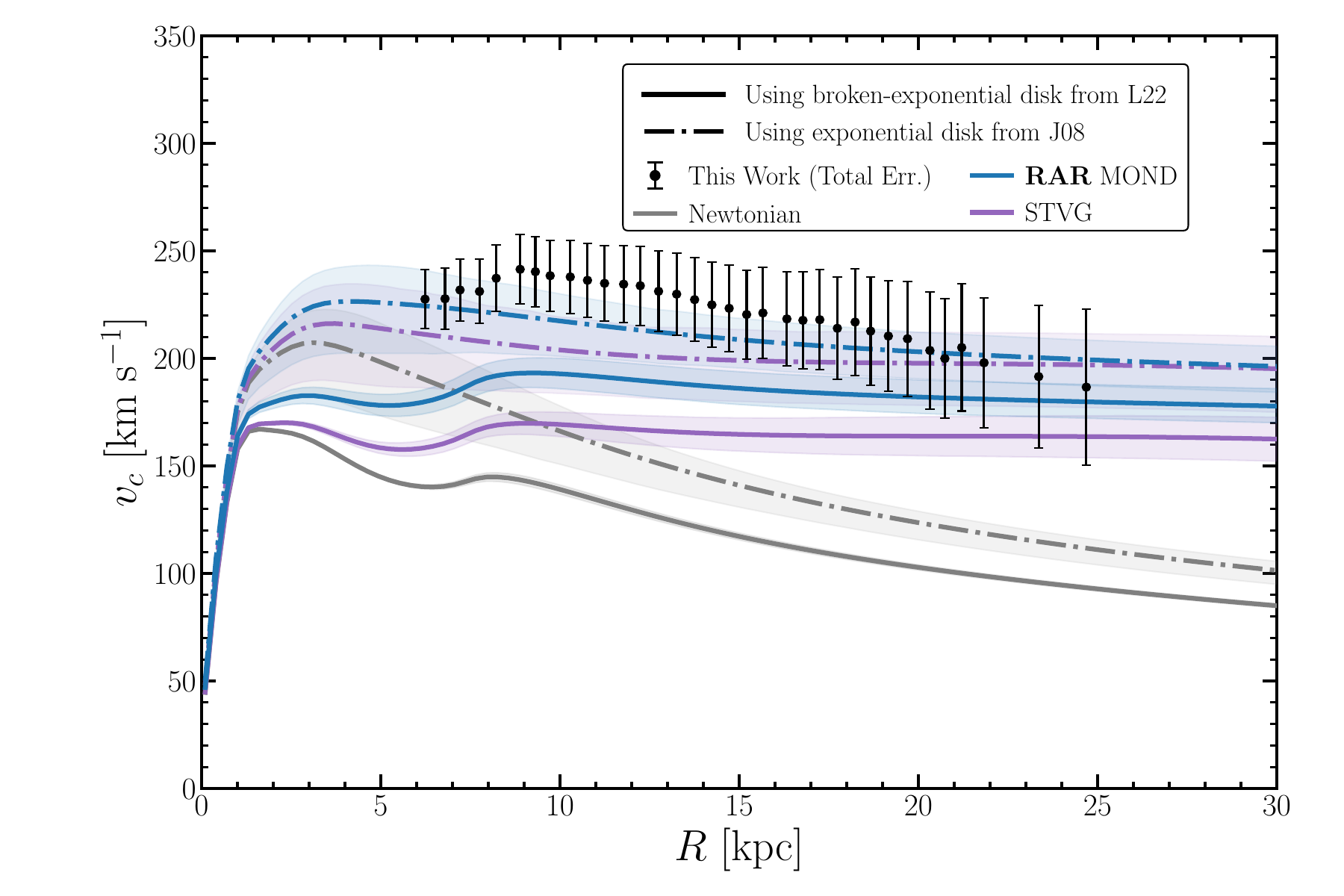} 
	\caption{\textbf{The modified gravity predicted rotation curves of the MW for previous and current stellar disk models.}
		The characteristic acceleration in MOND is fixed at the universal value $a_0 = 1.20 \pm 0.24 \times 10^{-10} \, \mathrm{m\,s^{-2}}$~\cite{McGaugh_2016}, and the STVG model adopts standard universal parameters ($\alpha = 8.89 \pm 0.34$, $\mu = 0.042 \pm 0.004 \, \mathrm{kpc^{-1}}$)~\cite{Moffat_2013}. We compare the broken-exponential disk from Lian et al. (\cite{Lian_2022}; hereafter L22) adopted in this work against the exponential disk from Juri\'c et al. (\cite{Juri__2008}; hereafter J08), with both models normalized to the same local density. The modified gravity models, in particular the MOND, can reasonably reproduce the observed rotation curves in the J08 disk scenario. This is however not the case for the L22 disk model since the signification reduction in baryonic mass leads to a systematic underprediction of velocities, resulting in a serious tension between the modified gravity predictions and the observations.}
	\label{fig:fitting_plot} 
\end{figure}

Specifically, while the local baryonic density in modified gravity models explains the vertical potential $\Phi_{z}$, 
albeit less successfully than the dark matter halo model~\cite{ Lisanti_2019, Davari_2020, Zhu_2022}, 
the radial acceleration from the broken‑exponential disk profile inferred from MAPs~\cite{Lian_2022} is insufficient to match the observed rotation curve.
Using a broken-exponential disk model, we find that 
the baryonic contribution to the MW rotation curve is substantially lower than inferred from traditional single-exponential fits (see Fig.~\ref{fig:fitting_plot}).
Note that for the MOND model, we only present results using the RAR interpolation function, which was identified as the optimal formulation prior to this study~\cite{Banik_2022}.
As a result, tuning parameters in modified gravity theories to enhance the rotation curve prediction inevitably undermines their ability to predict the vertical force, 
rendering it impossible to simultaneously fit both 
$v_{\rm c}$ and 
$\Phi_{z}$.

We examine the structural properties of the inferred dark matter halos to evaluate the robustness of the dark matter framework.
We find that the concentration parameters $c_{200}$~\cite{Dutton_2014}, derived from standard NFW profiles deviate significantly from dark-matter-only (DMO) cosmological expectations ($c_{200} \approx 27$ versus $\sim 10$~\cite{Maccio_2008}), yet align with models incorporating baryonic contraction~\cite{DiCintio_2014, Freundlich_2020}. 
In contrast, the $c_{200}$ values derived from the Einasto profile ($\approx 12.5$) fall entirely within the expected distribution of DMO simulations \cite{Dutton_2014}. A detailed investigation of these halo properties and their underlying physics will be presented elsewhere.

\begin{figure} 
	\centering
	\includegraphics[width=\linewidth]{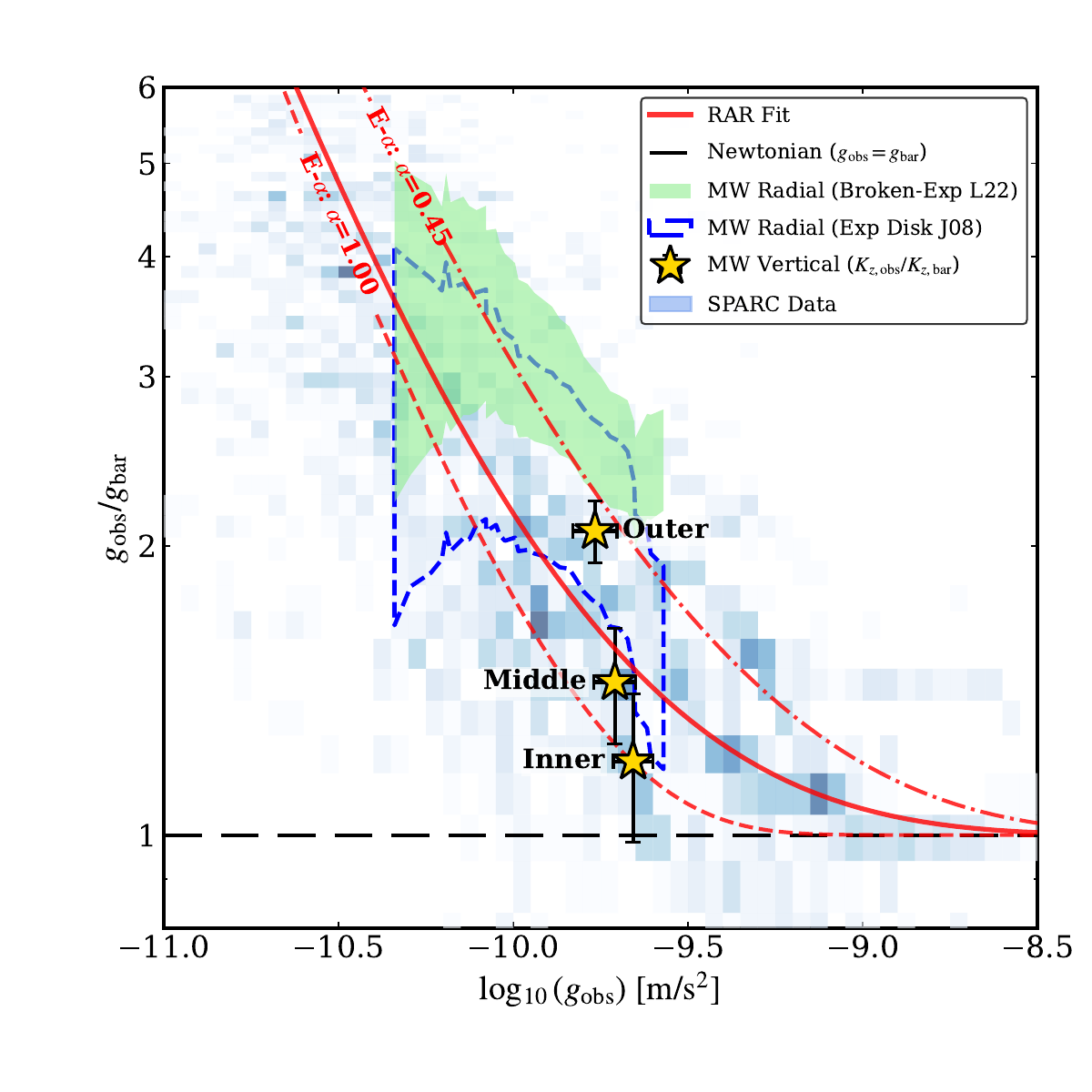} 
	\caption{\textbf{MW dynamical data in acceleration space.}
		Blue histogram shows the SPARC dataset, with the solid red line indicating the RAR fitting curve ~\cite{McGaugh_2016}. Red dashed and dash-dotted curves depict the $E$-$\alpha$ interpolation models with $\alpha=1.00$ and $\alpha=0.45$. The vertical axis plots the logarithmic ratio of observed-to-baryonic acceleration ($g_{\mathrm{obs}}/g_{\mathrm{bar}}$ or $K_{z,\mathrm{obs}}/K_{z,\mathrm{bar}}$). Here, $g_{\mathrm{obs}}$ is derived from the MW rotation curve constructed in this work, accounting for total error. For consistency, the gravitational contributions from the gas and bulge components are identical to the models used in our main fitting procedure. The light green shaded band denotes the $1\sigma$ radial enhancement from our broken-exponential disk model (L22), while the blue dashed contour marks the traditional single-exponential disk (J08). Gold star symbols represent vertical force enhancement constrained by the phase-space snail, derived via differencing the reconstructed potential $\Phi_z$. To reduce noise, data are binned into three radial regions (Inner, Middle, and Outer), each averaging three adjacent $R_{\mathrm{pot}}$ bins. }
	\label{fig:rar_projection} 
\end{figure}

To explore the failure of MOND, we project MW dynamical data into acceleration space (Fig.~\ref{fig:rar_projection}). The vertical axis $g_{\mathrm{obs}}/g_{\mathrm{bar}}$ represents the ratio of total kinematic acceleration to the Newtonian acceleration expected from baryonic matter alone, allowing a direct comparison with the vertical force ratio $K_{z, \mathrm{obs}}/K_{z, \mathrm{bar}}$ (golden stars). These quantities are approximately related as
\begin{equation}
\frac{K_{z, \mathrm{obs}}}{K_{z, \mathrm{bar}}} \approx \frac{g_{\mathrm{obs}}}{g_{\mathrm{bar}}} \approx \nu(g_N/a_0),
\end{equation}
Within the framework of modified gravity, all dynamical trajectories strictly follow a single, universal acceleration law dictated by the baryonic distribution.

Fig.~\ref{fig:rar_projection} reveals two fundamental inconsistencies in the MONDian framework. First, adopting the broken-exponential disk (L22) reduces the baryonic contribution relative to traditional models (J08), causing observed accelerations to exceed predictions, whereas J08 appears consistent with the universal RAR fit. Second, vertical forces derived from the phase-space snail decouple radially from the rotation data, exceeding RAR expectations at small radii while showing a deficit at large radii. 
To address these discrepancies, we attempt to reconcile the MOND theory by tailoring the $E$-$\alpha$ family (see Supplementary Material), where the parameter $\alpha$ controls the shape of the interpolation function. Our analysis reveals that $\alpha=0.66$ is comparable to the unsuccessful RAR projection. Furthermore, we find that $\alpha=0.45$ aligns only with the rotation curve data, while $\alpha=1.0$ fits only individual vertical observation points. 
Still, the $E$-$\alpha$ family fails to consistently explain observations. We have also studied even more contrived $\nu(y)$-functions with a sharp feature at $g_N \sim a_0$ \cite{Milgrom_2008} in order to mimic the fall of the vertical force ratio (cf. Fig.~\ref{fig:rar_projection}). However, the inconsistency of these contrived $\nu(y)$-functions with the Galactic radial force and extragalactic RAR is not resolved. Ultimately, no modified gravity model varying monotonically with $g_{\mathrm{bar}}$ can self-consistently bridge the rotation curve, vertical oscillation, and extragalactic RAR constraints with a single universal curve.

In contrast, by introducing a nearly spherical cold dark matter halo, dark matter models break the degeneracy between radial gravity and local matter density. 
The halo supplies substantial radial gravity in the outer MW disk while leaving the vertical potential near the Sun essentially unchanged, thereby satisfying both dynamical constraints self-consistently. 
Hence, the missing mass distribution inferred from the data is more naturally explained by a quasi-spherical dark matter halo than by a modified gravitational field rigidly tied to the baryonic disk.

To specifically test the robustness of the MOND framework, we performed a complementary analysis treating the characteristic acceleration $a_0$ as a free parameter. Even with this added flexibility, QUMOND remain statistically excluded at a $13$--$14\sigma$ significance level compared to the CDM paradigm. The best-fit $a_0$ for QUMOND ($1.5\times10^{-10}$ to $2.3\times10^{-10}~\mathrm{m\,s^{-2}}$) exceeds the universal value, 
while the implied thin-disk density ($\rho_{\odot,\mathrm{thin}} \approx 1.9\times10^{-2}$ to $2.4\times10^{-2}~\mathrm{M_\odot\,pc^{-3}}$) falls below empirical estimates. 
This structural incompatibility persists even when accounting for the External Field Effect, 
which only worsen the outer rotation curve discrepancy.

As a complementary test, we explore baryonic uncertainties by varying the thin stellar disk break radius $R_b$ and the HI gas parameters (normalization and $R_m$), which are detailed in the Supplementary Material. These components constitute the majority of the MW stellar and gaseous mass, respectively. Even with this added flexibility, modified gravity remains statistically disfavored relative to CDM.
For fitting to the rotation curve, the models drive $R_b$ toward zero, effectively reverting to a high-mass single-exponential disk to maximize stellar contribution. 
Conversely, to satisfy vertical potential constraints, they favor low gas masses and larger $R_m$ values. 
This irreconcilable tension with multi-wavelength data thus confirms a structural failure rather than a baryonic modeling artifact. 

In summary, modified gravity theories face a fundamental dual crisis in accounting for MW dynamics.
Statistically, they are decisively ruled out with overwhelming significance relative to the CDM framework ($>13\sigma$ for MOND and $>4.2\sigma$ for STVG).
Our results show that current modified gravity frameworks encounter considerable challenges simultaneously accommodating the Milky Way radial rotation curve, the vertical phase-space spirals revealed by Gaia, and the broken-exponential profile of the stellar disk.
Nonetheless, a complete picture awaits fuller inner MW data, which may be released in the future.

\textit{Acknowledgments.---}
This work was supported in part by the National Natural Science Foundation of China (NSFC; Grants No. 12588101 and No. 12588202), the National Key R\&D Program of China (Grants No. 2022YFF0503304, No. 2024YFA1611902, and No. 2024YFA1611903), the CAS Project for Young Scientists in Basic Research (Grants No. YSBR-092 and No. YSBR-062), and the China Manned Space Program (CMS; Grants No. CMS-CSST-2025-A03 and No. CMS-CSST-2025-A11).
Y.Z.F. acknowledges support from the New Cornerstone Science Foundation through the XPLORER PRIZE.
L.Z., X.-X.X., and H.-N.L. acknowledge support from the National Key R\&D Program of China and the CAS Project for Young Scientists in Basic Research.
H.-N.L. and X.-X.X. also acknowledge support from the Strategic Priority Research Program of CAS (Grants No. XDB1160100 and No. XDB1160102).
This work has made use of data from the European Space Agency (ESA) mission Gaia, processed by the Gaia Data Processing and Analysis Consortium (DPAC). We also acknowledge the use of data from SDSS-IV, 2MASS, and WISE.

\bibliographystyle{apsrev4-2}
\setcitestyle{numbers,sort&compress}
\bibliography{references}

\begin{thebibliography}{72}%
\makeatletter
\providecommand \@ifxundefined [1]{%
 \@ifx{#1\undefined}
}%
\providecommand \@ifnum [1]{%
 \ifnum #1\expandafter \@firstoftwo
 \else \expandafter \@secondoftwo
 \fi
}%
\providecommand \@ifx [1]{%
 \ifx #1\expandafter \@firstoftwo
 \else \expandafter \@secondoftwo
 \fi
}%
\providecommand \natexlab [1]{#1}%
\providecommand \enquote  [1]{``#1''}%
\providecommand \bibnamefont  [1]{#1}%
\providecommand \bibfnamefont [1]{#1}%
\providecommand \citenamefont [1]{#1}%
\providecommand \href@noop [0]{\@secondoftwo}%
\providecommand \href [0]{\begingroup \@sanitize@url \@href}%
\providecommand \@href[1]{\@@startlink{#1}\@@href}%
\providecommand \@@href[1]{\endgroup#1\@@endlink}%
\providecommand \@sanitize@url [0]{\catcode `\\12\catcode `\$12\catcode
  `\&12\catcode `\#12\catcode `\^12\catcode `\_12\catcode `\%12\relax}%
\providecommand \@@startlink[1]{}%
\providecommand \@@endlink[0]{}%
\providecommand \url  [0]{\begingroup\@sanitize@url \@url }%
\providecommand \@url [1]{\endgroup\@href {#1}{\urlprefix }}%
\providecommand \urlprefix  [0]{URL }%
\providecommand \Eprint [0]{\href }%
\providecommand \doibase [0]{https://doi.org/}%
\providecommand \selectlanguage [0]{\@gobble}%
\providecommand \bibinfo  [0]{\@secondoftwo}%
\providecommand \bibfield  [0]{\@secondoftwo}%
\providecommand \translation [1]{[#1]}%
\providecommand \BibitemOpen [0]{}%
\providecommand \bibitemStop [0]{}%
\providecommand \bibitemNoStop [0]{.\EOS\space}%
\providecommand \EOS [0]{\spacefactor3000\relax}%
\providecommand \BibitemShut  [1]{\csname bibitem#1\endcsname}%
\let\auto@bib@innerbib\@empty
\bibitem [{\citenamefont {Bullock}\ and\ \citenamefont
  {Boylan-Kolchin}(2017)}]{Bullock_2017}%
  \BibitemOpen
  \bibfield  {author} {\bibinfo {author} {\bibfnamefont {J.~S.}\ \bibnamefont
  {Bullock}}\ and\ \bibinfo {author} {\bibfnamefont {M.}~\bibnamefont
  {Boylan-Kolchin}},\ }\href
  {https://doi.org/10.1146/annurev-astro-091916-055313} {\bibfield  {journal}
  {\bibinfo  {journal} {Annual Review of Astronomy and Astrophysics}\ }\textbf
  {\bibinfo {volume} {55}},\ \bibinfo {pages} {343–387} (\bibinfo {year}
  {2017})}\BibitemShut {NoStop}%
\bibitem [{\citenamefont {Banik}\ and\ \citenamefont
  {Zhao}(2022)}]{Banik_2022}%
  \BibitemOpen
  \bibfield  {author} {\bibinfo {author} {\bibfnamefont {I.}~\bibnamefont
  {Banik}}\ and\ \bibinfo {author} {\bibfnamefont {H.}~\bibnamefont {Zhao}},\
  }\href {https://doi.org/10.3390/sym14071331} {\bibfield  {journal} {\bibinfo
  {journal} {Symmetry}\ }\textbf {\bibinfo {volume} {14}},\ \bibinfo {pages}
  {1331} (\bibinfo {year} {2022})}\BibitemShut {NoStop}%
\bibitem [{\citenamefont {{Milgrom}}(1983)}]{Milgrom_1983}%
  \BibitemOpen
  \bibfield  {author} {\bibinfo {author} {\bibfnamefont {M.}~\bibnamefont
  {{Milgrom}}},\ }\href {https://doi.org/10.1086/161130} {\bibfield  {journal}
  {\bibinfo  {journal} {The Astrophysical Journal}\ }\textbf {\bibinfo {volume}
  {270}},\ \bibinfo {pages} {365} (\bibinfo {year} {1983})}\BibitemShut
  {NoStop}%
\bibitem [{\citenamefont {{Famaey}}\ and\ \citenamefont
  {{McGaugh}}(2012)}]{Famaey_2012}%
  \BibitemOpen
  \bibfield  {author} {\bibinfo {author} {\bibfnamefont {B.}~\bibnamefont
  {{Famaey}}}\ and\ \bibinfo {author} {\bibfnamefont {S.~S.}\ \bibnamefont
  {{McGaugh}}},\ }\href {https://doi.org/10.12942/lrr-2012-10} {\bibfield
  {journal} {\bibinfo  {journal} {Living Reviews in Relativity}\ }\textbf
  {\bibinfo {volume} {15}},\ \bibinfo {eid} {10} (\bibinfo {year} {2012})},\
  \Eprint {https://arxiv.org/abs/1112.3960} {arXiv:1112.3960 [astro-ph.CO]}
  \BibitemShut {NoStop}%
\bibitem [{\citenamefont {{McGaugh}}\ \emph {et~al.}(2016)\citenamefont
  {{McGaugh}}, \citenamefont {{Lelli}},\ and\ \citenamefont
  {{Schombert}}}]{McGaugh_2016}%
  \BibitemOpen
  \bibfield  {author} {\bibinfo {author} {\bibfnamefont {S.~S.}\ \bibnamefont
  {{McGaugh}}}, \bibinfo {author} {\bibfnamefont {F.}~\bibnamefont {{Lelli}}},\
  and\ \bibinfo {author} {\bibfnamefont {J.~M.}\ \bibnamefont {{Schombert}}},\
  }\href {https://doi.org/10.1103/PhysRevLett.117.201101} {\bibfield  {journal}
  {\bibinfo  {journal} {Physical Review Letters}\ }\textbf {\bibinfo {volume}
  {117}},\ \bibinfo {eid} {201101} (\bibinfo {year} {2016})},\ \Eprint
  {https://arxiv.org/abs/1609.05917} {arXiv:1609.05917 [astro-ph.GA]}
  \BibitemShut {NoStop}%
\bibitem [{\citenamefont {McGaugh}(2020)}]{McGaugh_2020}%
  \BibitemOpen
  \bibfield  {author} {\bibinfo {author} {\bibfnamefont {S.}~\bibnamefont
  {McGaugh}},\ }\href {https://doi.org/10.3390/galaxies8020035} {\bibfield
  {journal} {\bibinfo  {journal} {Galaxies}\ }\textbf {\bibinfo {volume} {8}},\
  \bibinfo {pages} {35} (\bibinfo {year} {2020})}\BibitemShut {NoStop}%
\bibitem [{\citenamefont {Roshan}\ \emph {et~al.}(2021)\citenamefont {Roshan},
  \citenamefont {Banik}, \citenamefont {Ghafourian}, \citenamefont {Thies},
  \citenamefont {Famaey}, \citenamefont {Asencio},\ and\ \citenamefont
  {Kroupa}}]{Roshan_2021}%
  \BibitemOpen
  \bibfield  {author} {\bibinfo {author} {\bibfnamefont {M.}~\bibnamefont
  {Roshan}}, \bibinfo {author} {\bibfnamefont {I.}~\bibnamefont {Banik}},
  \bibinfo {author} {\bibfnamefont {N.}~\bibnamefont {Ghafourian}}, \bibinfo
  {author} {\bibfnamefont {I.}~\bibnamefont {Thies}}, \bibinfo {author}
  {\bibfnamefont {B.}~\bibnamefont {Famaey}}, \bibinfo {author} {\bibfnamefont
  {E.}~\bibnamefont {Asencio}},\ and\ \bibinfo {author} {\bibfnamefont
  {P.}~\bibnamefont {Kroupa}},\ }\href {https://doi.org/10.1093/mnras/stab651}
  {\bibfield  {journal} {\bibinfo  {journal} {Monthly Notices of the Royal
  Astronomical Society}\ }\textbf {\bibinfo {volume} {503}},\ \bibinfo {pages}
  {2833–2860} (\bibinfo {year} {2021})}\BibitemShut {NoStop}%
\bibitem [{\citenamefont {Moffat}\ and\ \citenamefont
  {Rahvar}(2013)}]{Moffat_2013}%
  \BibitemOpen
  \bibfield  {author} {\bibinfo {author} {\bibfnamefont {J.~W.}\ \bibnamefont
  {Moffat}}\ and\ \bibinfo {author} {\bibfnamefont {S.}~\bibnamefont
  {Rahvar}},\ }\href {https://doi.org/10.1093/mnras/stt1670} {\bibfield
  {journal} {\bibinfo  {journal} {Monthly Notices of the Royal Astronomical
  Society}\ }\textbf {\bibinfo {volume} {436}},\ \bibinfo {pages} {1439}
  (\bibinfo {year} {2013})}\BibitemShut {NoStop}%
\bibitem [{\citenamefont {Green}\ and\ \citenamefont
  {Moffat}(2019)}]{Green_2019}%
  \BibitemOpen
  \bibfield  {author} {\bibinfo {author} {\bibfnamefont {M.}~\bibnamefont
  {Green}}\ and\ \bibinfo {author} {\bibfnamefont {J.}~\bibnamefont {Moffat}},\
  }\href {https://doi.org/10.1016/j.dark.2019.100323} {\bibfield  {journal}
  {\bibinfo  {journal} {Physics of the Dark Universe}\ }\textbf {\bibinfo
  {volume} {25}},\ \bibinfo {pages} {100323} (\bibinfo {year}
  {2019})}\BibitemShut {NoStop}%
\bibitem [{\citenamefont {Chae}(2023)}]{Chae:2023prf}%
  \BibitemOpen
  \bibfield  {author} {\bibinfo {author} {\bibfnamefont {K.-H.}\ \bibnamefont
  {Chae}},\ }\href {https://doi.org/10.3847/1538-4357/ace101} {\bibfield
  {journal} {\bibinfo  {journal} {Astrophys. J.}\ }\textbf {\bibinfo {volume}
  {952}},\ \bibinfo {pages} {128} (\bibinfo {year} {2023})},\ \bibinfo {note}
  {[Erratum: Astrophys.J. 956, 69 (2023)]},\ \Eprint
  {https://arxiv.org/abs/2305.04613} {arXiv:2305.04613 [astro-ph.GA]}
  \BibitemShut {NoStop}%
\bibitem [{\citenamefont {Chae}\ \emph {et~al.}(2026)\citenamefont {Chae},
  \citenamefont {Lee}, \citenamefont {Hernandez}, \citenamefont {Orlov},
  \citenamefont {Lim}, \citenamefont {Turnshek},\ and\ \citenamefont
  {Lee}}]{chae2026}%
  \BibitemOpen
  \bibfield  {author} {\bibinfo {author} {\bibfnamefont {K.~H.}\ \bibnamefont
  {Chae}}, \bibinfo {author} {\bibfnamefont {B.~C.}\ \bibnamefont {Lee}},
  \bibinfo {author} {\bibfnamefont {X.}~\bibnamefont {Hernandez}}, \bibinfo
  {author} {\bibfnamefont {V.~G.}\ \bibnamefont {Orlov}}, \bibinfo {author}
  {\bibfnamefont {D.}~\bibnamefont {Lim}}, \bibinfo {author} {\bibfnamefont
  {D.~A.}\ \bibnamefont {Turnshek}},\ and\ \bibinfo {author} {\bibfnamefont
  {Y.~W.}\ \bibnamefont {Lee}},\ }\href {https://arxiv.org/abs/2601.21728}
  {\bibinfo {title} {Detection of gravitational anomaly at low acceleration
  from a highest-quality sample of 36 wide binaries with accurate 3d
  velocities}} (\bibinfo {year} {2026}),\ \Eprint
  {https://arxiv.org/abs/2601.21728} {arXiv:2601.21728 [astro-ph.GA]}
  \BibitemShut {NoStop}%
\bibitem [{\citenamefont {{Haslbauer}}\ \emph {et~al.}(2020)\citenamefont
  {{Haslbauer}}, \citenamefont {{Banik}},\ and\ \citenamefont
  {{Kroupa}}}]{Haslbauer_2020}%
  \BibitemOpen
  \bibfield  {author} {\bibinfo {author} {\bibfnamefont {M.}~\bibnamefont
  {{Haslbauer}}}, \bibinfo {author} {\bibfnamefont {I.}~\bibnamefont
  {{Banik}}},\ and\ \bibinfo {author} {\bibfnamefont {P.}~\bibnamefont
  {{Kroupa}}},\ }\href {https://doi.org/10.1093/mnras/staa2348} {\bibfield
  {journal} {\bibinfo  {journal} {Monthly Notices of the Royal Astronomical
  Society}\ }\textbf {\bibinfo {volume} {499}},\ \bibinfo {pages} {2845}
  (\bibinfo {year} {2020})},\ \Eprint {https://arxiv.org/abs/2009.11292}
  {arXiv:2009.11292 [astro-ph.CO]} \BibitemShut {NoStop}%
\bibitem [{\citenamefont {Angus}(2009)}]{Angus_2009}%
  \BibitemOpen
  \bibfield  {author} {\bibinfo {author} {\bibfnamefont {G.~W.}\ \bibnamefont
  {Angus}},\ }\href {https://doi.org/10.1111/j.1365-2966.2008.14341.x}
  {\bibfield  {journal} {\bibinfo  {journal} {Monthly Notices of the Royal
  Astronomical Society}\ }\textbf {\bibinfo {volume} {394}},\ \bibinfo {pages}
  {527–532} (\bibinfo {year} {2009})}\BibitemShut {NoStop}%
\bibitem [{\citenamefont {B{\'i}lek}\ and\ \citenamefont
  {Zhao}(2025)}]{Bilek_2025}%
  \BibitemOpen
  \bibfield  {author} {\bibinfo {author} {\bibfnamefont {M.}~\bibnamefont
  {B{\'i}lek}}\ and\ \bibinfo {author} {\bibfnamefont {H.}~\bibnamefont
  {Zhao}},\ }\bibfield  {journal} {\bibinfo  {journal} {Astronomy \&
  Astrophysics}\ }\textbf {\bibinfo {volume} {698}},\ \href
  {https://doi.org/10.1051/0004-6361/202452250} {10.1051/0004-6361/202452250}
  (\bibinfo {year} {2025})\BibitemShut {NoStop}%
\bibitem [{\citenamefont {Russell}\ \emph {et~al.}(2026)\citenamefont
  {Russell}, \citenamefont {Banik}, \citenamefont {Cray},\ and\ \citenamefont
  {Zhao}}]{Russell_2026}%
  \BibitemOpen
  \bibfield  {author} {\bibinfo {author} {\bibfnamefont {A.}~\bibnamefont
  {Russell}}, \bibinfo {author} {\bibfnamefont {I.}~\bibnamefont {Banik}},
  \bibinfo {author} {\bibfnamefont {O.}~\bibnamefont {Cray}},\ and\ \bibinfo
  {author} {\bibfnamefont {H.}~\bibnamefont {Zhao}},\ }\bibfield  {journal}
  {\bibinfo  {journal} {Monthly Notices of the Royal Astronomical Society}\
  }\href {https://doi.org/10.1093/mnras/stag399} {10.1093/mnras/stag399}
  (\bibinfo {year} {2026})\BibitemShut {NoStop}%
\bibitem [{\citenamefont {Haslbauer}\ \emph {et~al.}(2024)\citenamefont
  {Haslbauer}, \citenamefont {Banik}, \citenamefont {Kroupa}, \citenamefont
  {Zhao},\ and\ \citenamefont {Asencio}}]{Haslbauer_2024}%
  \BibitemOpen
  \bibfield  {author} {\bibinfo {author} {\bibfnamefont {M.}~\bibnamefont
  {Haslbauer}}, \bibinfo {author} {\bibfnamefont {I.}~\bibnamefont {Banik}},
  \bibinfo {author} {\bibfnamefont {P.}~\bibnamefont {Kroupa}}, \bibinfo
  {author} {\bibfnamefont {H.}~\bibnamefont {Zhao}},\ and\ \bibinfo {author}
  {\bibfnamefont {E.}~\bibnamefont {Asencio}},\ }\href
  {https://doi.org/10.3390/universe10100385} {\bibfield  {journal} {\bibinfo
  {journal} {Universe}\ }\textbf {\bibinfo {volume} {10}},\ \bibinfo {pages}
  {385} (\bibinfo {year} {2024})}\BibitemShut {NoStop}%
\bibitem [{\citenamefont {Banik}\ \emph {et~al.}(2024)\citenamefont {Banik},
  \citenamefont {Pittordis}, \citenamefont {Sutherland}, \citenamefont
  {Famaey}, \citenamefont {Ibata}, \citenamefont {Mieske},\ and\ \citenamefont
  {Zhao}}]{Banik:2023pbo}%
  \BibitemOpen
  \bibfield  {author} {\bibinfo {author} {\bibfnamefont {I.}~\bibnamefont
  {Banik}}, \bibinfo {author} {\bibfnamefont {C.}~\bibnamefont {Pittordis}},
  \bibinfo {author} {\bibfnamefont {W.}~\bibnamefont {Sutherland}}, \bibinfo
  {author} {\bibfnamefont {B.}~\bibnamefont {Famaey}}, \bibinfo {author}
  {\bibfnamefont {R.}~\bibnamefont {Ibata}}, \bibinfo {author} {\bibfnamefont
  {S.}~\bibnamefont {Mieske}},\ and\ \bibinfo {author} {\bibfnamefont
  {H.}~\bibnamefont {Zhao}},\ }\href {https://doi.org/10.1093/mnras/stad3393}
  {\bibfield  {journal} {\bibinfo  {journal} {Mon. Not. Roy. Astron. Soc.}\
  }\textbf {\bibinfo {volume} {527}},\ \bibinfo {pages} {4573} (\bibinfo {year}
  {2024})},\ \Eprint {https://arxiv.org/abs/2311.03436} {arXiv:2311.03436
  [astro-ph.SR]} \BibitemShut {NoStop}%
\bibitem [{\citenamefont {Alam}\ \emph {et~al.}(2017)\citenamefont {Alam},
  \citenamefont {Ata}, \citenamefont {Bailey} \emph {et~al.}}]{Alam_2017}%
  \BibitemOpen
  \bibfield  {author} {\bibinfo {author} {\bibfnamefont {S.}~\bibnamefont
  {Alam}}, \bibinfo {author} {\bibfnamefont {M.}~\bibnamefont {Ata}}, \bibinfo
  {author} {\bibfnamefont {S.}~\bibnamefont {Bailey}}, \emph {et~al.},\ }\href
  {https://doi.org/10.1093/mnras/stx721} {\bibfield  {journal} {\bibinfo
  {journal} {Monthly Notices of the Royal Astronomical Society}\ }\textbf
  {\bibinfo {volume} {470}},\ \bibinfo {pages} {2617} (\bibinfo {year}
  {2017})}\BibitemShut {NoStop}%
\bibitem [{\citenamefont {Aghanim}\ \emph {et~al.}(2020)\citenamefont
  {Aghanim}, \citenamefont {Akrami}, \citenamefont {Ashdown} \emph
  {et~al.}}]{Aghanim_2020}%
  \BibitemOpen
  \bibfield  {author} {\bibinfo {author} {\bibfnamefont {N.}~\bibnamefont
  {Aghanim}}, \bibinfo {author} {\bibfnamefont {Y.}~\bibnamefont {Akrami}},
  \bibinfo {author} {\bibfnamefont {M.}~\bibnamefont {Ashdown}}, \emph
  {et~al.},\ }\href {https://doi.org/10.1051/0004-6361/201833910} {\bibfield
  {journal} {\bibinfo  {journal} {Astronomy \& Astrophysics}\ }\textbf
  {\bibinfo {volume} {641}},\ \bibinfo {pages} {A6} (\bibinfo {year}
  {2020})}\BibitemShut {NoStop}%
\bibitem [{\citenamefont {Juri{\'c}}\ \emph {et~al.}(2008)\citenamefont
  {Juri{\'c}} \emph {et~al.}}]{Juri__2008}%
  \BibitemOpen
  \bibfield  {author} {\bibinfo {author} {\bibfnamefont {M.}~\bibnamefont
  {Juri{\'c}}} \emph {et~al.},\ }\href {https://doi.org/10.1086/523619}
  {\bibfield  {journal} {\bibinfo  {journal} {The Astrophysical Journal}\
  }\textbf {\bibinfo {volume} {673}},\ \bibinfo {pages} {864} (\bibinfo {year}
  {2008})}\BibitemShut {NoStop}%
\bibitem [{\citenamefont {Nipoti}\ \emph {et~al.}(2007)\citenamefont {Nipoti},
  \citenamefont {Londrillo}, \citenamefont {Zhao},\ and\ \citenamefont
  {Ciotti}}]{Nipoti_2007}%
  \BibitemOpen
  \bibfield  {author} {\bibinfo {author} {\bibfnamefont {C.}~\bibnamefont
  {Nipoti}}, \bibinfo {author} {\bibfnamefont {P.}~\bibnamefont {Londrillo}},
  \bibinfo {author} {\bibfnamefont {H.~S.}\ \bibnamefont {Zhao}},\ and\
  \bibinfo {author} {\bibfnamefont {L.}~\bibnamefont {Ciotti}},\ }\href
  {https://doi.org/10.1111/j.1365-2966.2007.11835.x} {\bibfield  {journal}
  {\bibinfo  {journal} {Monthly Notices of the Royal Astronomical Society}\
  }\textbf {\bibinfo {volume} {379}},\ \bibinfo {pages} {597} (\bibinfo {year}
  {2007})}\BibitemShut {NoStop}%
\bibitem [{\citenamefont {{Lisanti}}\ \emph {et~al.}(2019)\citenamefont
  {{Lisanti}}, \citenamefont {{Moschella}}, \citenamefont {{Outmezguine}},\
  and\ \citenamefont {{Slone}}}]{Lisanti_2019}%
  \BibitemOpen
  \bibfield  {author} {\bibinfo {author} {\bibfnamefont {M.}~\bibnamefont
  {{Lisanti}}}, \bibinfo {author} {\bibfnamefont {M.}~\bibnamefont
  {{Moschella}}}, \bibinfo {author} {\bibfnamefont {N.~J.}\ \bibnamefont
  {{Outmezguine}}},\ and\ \bibinfo {author} {\bibfnamefont {O.}~\bibnamefont
  {{Slone}}},\ }\href {https://doi.org/10.1103/PhysRevD.100.083009} {\bibfield
  {journal} {\bibinfo  {journal} {Physical Review D}\ }\textbf {\bibinfo
  {volume} {100}},\ \bibinfo {eid} {083009} (\bibinfo {year} {2019})},\ \Eprint
  {https://arxiv.org/abs/1812.08169} {arXiv:1812.08169 [astro-ph.GA]}
  \BibitemShut {NoStop}%
\bibitem [{\citenamefont {Davari}\ and\ \citenamefont
  {Rahvar}(2020)}]{Davari_2020}%
  \BibitemOpen
  \bibfield  {author} {\bibinfo {author} {\bibfnamefont {Z.}~\bibnamefont
  {Davari}}\ and\ \bibinfo {author} {\bibfnamefont {S.}~\bibnamefont
  {Rahvar}},\ }\href {https://doi.org/10.1093/mnras/staa1660} {\bibfield
  {journal} {\bibinfo  {journal} {Monthly Notices of the Royal Astronomical
  Society}\ }\textbf {\bibinfo {volume} {496}},\ \bibinfo {pages} {3502}
  (\bibinfo {year} {2020})}\BibitemShut {NoStop}%
\bibitem [{\citenamefont {Zhu}\ \emph {et~al.}(2022)\citenamefont {Zhu},
  \citenamefont {Ma}, \citenamefont {Dong}, \citenamefont {Huang},
  \citenamefont {Mistele}, \citenamefont {Peng}, \citenamefont {Long},
  \citenamefont {Wang}, \citenamefont {Chang},\ and\ \citenamefont
  {Jin}}]{Zhu_2022}%
  \BibitemOpen
  \bibfield  {author} {\bibinfo {author} {\bibfnamefont {Y.}~\bibnamefont
  {Zhu}}, \bibinfo {author} {\bibfnamefont {H.-X.}\ \bibnamefont {Ma}},
  \bibinfo {author} {\bibfnamefont {X.-B.}\ \bibnamefont {Dong}}, \bibinfo
  {author} {\bibfnamefont {Y.}~\bibnamefont {Huang}}, \bibinfo {author}
  {\bibfnamefont {T.}~\bibnamefont {Mistele}}, \bibinfo {author} {\bibfnamefont
  {B.}~\bibnamefont {Peng}}, \bibinfo {author} {\bibfnamefont {Q.}~\bibnamefont
  {Long}}, \bibinfo {author} {\bibfnamefont {T.}~\bibnamefont {Wang}}, \bibinfo
  {author} {\bibfnamefont {L.}~\bibnamefont {Chang}},\ and\ \bibinfo {author}
  {\bibfnamefont {X.}~\bibnamefont {Jin}},\ }\href
  {https://doi.org/10.1093/mnras/stac3483} {\bibfield  {journal} {\bibinfo
  {journal} {Monthly Notices of the Royal Astronomical Society}\ }\textbf
  {\bibinfo {volume} {519}},\ \bibinfo {pages} {4479} (\bibinfo {year}
  {2022})}\BibitemShut {NoStop}%
\bibitem [{\citenamefont {{L{\'o}pez-Corredoira}}(2025)}]{Lopez_2025}%
  \BibitemOpen
  \bibfield  {author} {\bibinfo {author} {\bibfnamefont {M.}~\bibnamefont
  {{L{\'o}pez-Corredoira}}},\ }\href {https://doi.org/10.3847/1538-4357/ad94f5}
  {\bibfield  {journal} {\bibinfo  {journal} {The Astrophysical Journal}\
  }\textbf {\bibinfo {volume} {978}},\ \bibinfo {eid} {45} (\bibinfo {year}
  {2025})},\ \Eprint {https://arxiv.org/abs/2412.09665} {arXiv:2412.09665
  [astro-ph.GA]} \BibitemShut {NoStop}%
\bibitem [{\citenamefont {Antoja}\ \emph {et~al.}(2018)\citenamefont {Antoja},
  \citenamefont {Helmi}, \citenamefont {Romero-Gómez}, \citenamefont {Katz},
  \citenamefont {Babusiaux}, \citenamefont {Drimmel}, \citenamefont {Evans},
  \citenamefont {Figueras}, \citenamefont {Poggio}, \citenamefont {Reylé},
  \citenamefont {Robin}, \citenamefont {Seabroke},\ and\ \citenamefont
  {Soubiran}}]{Antoja_2018}%
  \BibitemOpen
  \bibfield  {author} {\bibinfo {author} {\bibfnamefont {T.}~\bibnamefont
  {Antoja}}, \bibinfo {author} {\bibfnamefont {A.}~\bibnamefont {Helmi}},
  \bibinfo {author} {\bibfnamefont {M.}~\bibnamefont {Romero-Gómez}}, \bibinfo
  {author} {\bibfnamefont {D.}~\bibnamefont {Katz}}, \bibinfo {author}
  {\bibfnamefont {C.}~\bibnamefont {Babusiaux}}, \bibinfo {author}
  {\bibfnamefont {R.}~\bibnamefont {Drimmel}}, \bibinfo {author} {\bibfnamefont
  {D.~W.}\ \bibnamefont {Evans}}, \bibinfo {author} {\bibfnamefont
  {F.}~\bibnamefont {Figueras}}, \bibinfo {author} {\bibfnamefont
  {E.}~\bibnamefont {Poggio}}, \bibinfo {author} {\bibfnamefont
  {C.}~\bibnamefont {Reylé}}, \bibinfo {author} {\bibfnamefont {A.~C.}\
  \bibnamefont {Robin}}, \bibinfo {author} {\bibfnamefont {G.}~\bibnamefont
  {Seabroke}},\ and\ \bibinfo {author} {\bibfnamefont {C.}~\bibnamefont
  {Soubiran}},\ }\href {https://doi.org/10.1038/s41586-018-0510-7} {\bibfield
  {journal} {\bibinfo  {journal} {Nature}\ }\textbf {\bibinfo {volume} {561}},\
  \bibinfo {pages} {360} (\bibinfo {year} {2018})}\BibitemShut {NoStop}%
\bibitem [{\citenamefont {{Guo}}\ \emph {et~al.}(2024)\citenamefont {{Guo}},
  \citenamefont {{Li}}, \citenamefont {{Shen}}, \citenamefont {{Mao}},\ and\
  \citenamefont {{Liu}}}]{Guo_2024}%
  \BibitemOpen
  \bibfield  {author} {\bibinfo {author} {\bibfnamefont {R.}~\bibnamefont
  {{Guo}}}, \bibinfo {author} {\bibfnamefont {Z.-Y.}\ \bibnamefont {{Li}}},
  \bibinfo {author} {\bibfnamefont {J.}~\bibnamefont {{Shen}}}, \bibinfo
  {author} {\bibfnamefont {S.}~\bibnamefont {{Mao}}},\ and\ \bibinfo {author}
  {\bibfnamefont {C.}~\bibnamefont {{Liu}}},\ }\href
  {https://doi.org/10.3847/1538-4357/ad037b} {\bibfield  {journal} {\bibinfo
  {journal} {The Astrophysical Journal}\ }\textbf {\bibinfo {volume} {960}},\
  \bibinfo {eid} {133} (\bibinfo {year} {2024})},\ \Eprint
  {https://arxiv.org/abs/2310.10225} {arXiv:2310.10225 [astro-ph.GA]}
  \BibitemShut {NoStop}%
\bibitem [{\citenamefont {Bovy}\ \emph {et~al.}(2016)\citenamefont {Bovy},
  \citenamefont {Rix}, \citenamefont {Schlafly}, \citenamefont {Nidever},
  \citenamefont {Holtzman}, \citenamefont {Shetrone},\ and\ \citenamefont
  {Beers}}]{Bovy_2016}%
  \BibitemOpen
  \bibfield  {author} {\bibinfo {author} {\bibfnamefont {J.}~\bibnamefont
  {Bovy}}, \bibinfo {author} {\bibfnamefont {H.-W.}\ \bibnamefont {Rix}},
  \bibinfo {author} {\bibfnamefont {E.~F.}\ \bibnamefont {Schlafly}}, \bibinfo
  {author} {\bibfnamefont {D.~L.}\ \bibnamefont {Nidever}}, \bibinfo {author}
  {\bibfnamefont {J.~A.}\ \bibnamefont {Holtzman}}, \bibinfo {author}
  {\bibfnamefont {M.}~\bibnamefont {Shetrone}},\ and\ \bibinfo {author}
  {\bibfnamefont {T.~C.}\ \bibnamefont {Beers}},\ }\href
  {https://doi.org/10.3847/0004-637X/823/1/30} {\bibfield  {journal} {\bibinfo
  {journal} {The Astrophysical Journal}\ }\textbf {\bibinfo {volume} {823}},\
  \bibinfo {pages} {30} (\bibinfo {year} {2016})}\BibitemShut {NoStop}%
\bibitem [{\citenamefont {Ted~Mackereth}\ \emph {et~al.}(2017)\citenamefont
  {Ted~Mackereth}, \citenamefont {Bovy}, \citenamefont {Schiavon},
  \citenamefont {Zasowski}, \citenamefont {Cunha}, \citenamefont {Frinchaboy},
  \citenamefont {García~Perez}, \citenamefont {Hayden}, \citenamefont
  {Holtzman}, \citenamefont {Majewski}, \citenamefont {Mészáros},
  \citenamefont {Nidever}, \citenamefont {Pinsonneault},\ and\ \citenamefont
  {Shetrone}}]{Mackereth_2017}%
  \BibitemOpen
  \bibfield  {author} {\bibinfo {author} {\bibfnamefont {J.}~\bibnamefont
  {Ted~Mackereth}}, \bibinfo {author} {\bibfnamefont {J.}~\bibnamefont {Bovy}},
  \bibinfo {author} {\bibfnamefont {R.~P.}\ \bibnamefont {Schiavon}}, \bibinfo
  {author} {\bibfnamefont {G.}~\bibnamefont {Zasowski}}, \bibinfo {author}
  {\bibfnamefont {K.}~\bibnamefont {Cunha}}, \bibinfo {author} {\bibfnamefont
  {P.~M.}\ \bibnamefont {Frinchaboy}}, \bibinfo {author} {\bibfnamefont
  {A.~E.}\ \bibnamefont {García~Perez}}, \bibinfo {author} {\bibfnamefont
  {M.~R.}\ \bibnamefont {Hayden}}, \bibinfo {author} {\bibfnamefont
  {J.}~\bibnamefont {Holtzman}}, \bibinfo {author} {\bibfnamefont {S.~R.}\
  \bibnamefont {Majewski}}, \bibinfo {author} {\bibfnamefont {S.}~\bibnamefont
  {Mészáros}}, \bibinfo {author} {\bibfnamefont {D.~L.}\ \bibnamefont
  {Nidever}}, \bibinfo {author} {\bibfnamefont {M.}~\bibnamefont
  {Pinsonneault}},\ and\ \bibinfo {author} {\bibfnamefont {M.~D.}\ \bibnamefont
  {Shetrone}},\ }\href {https://doi.org/10.1093/mnras/stx1774} {\bibfield
  {journal} {\bibinfo  {journal} {Monthly Notices of the Royal Astronomical
  Society}\ }\textbf {\bibinfo {volume} {471}},\ \bibinfo {pages} {3057}
  (\bibinfo {year} {2017})}\BibitemShut {NoStop}%
\bibitem [{\citenamefont {Lian}\ \emph {et~al.}(2022)\citenamefont {Lian},
  \citenamefont {Zasowski}, \citenamefont {Mackereth}, \citenamefont {Imig},
  \citenamefont {Holtzman}, \citenamefont {Beaton}, \citenamefont {Bird},
  \citenamefont {Cunha}, \citenamefont {Fernández-Trincado}, \citenamefont
  {Horta}, \citenamefont {Lane}, \citenamefont {Masters}, \citenamefont
  {Nitschelm},\ and\ \citenamefont {Roman-Lopes}}]{Lian_2022}%
  \BibitemOpen
  \bibfield  {author} {\bibinfo {author} {\bibfnamefont {J.}~\bibnamefont
  {Lian}}, \bibinfo {author} {\bibfnamefont {G.}~\bibnamefont {Zasowski}},
  \bibinfo {author} {\bibfnamefont {T.}~\bibnamefont {Mackereth}}, \bibinfo
  {author} {\bibfnamefont {J.}~\bibnamefont {Imig}}, \bibinfo {author}
  {\bibfnamefont {J.~A.}\ \bibnamefont {Holtzman}}, \bibinfo {author}
  {\bibfnamefont {R.~L.}\ \bibnamefont {Beaton}}, \bibinfo {author}
  {\bibfnamefont {J.~C.}\ \bibnamefont {Bird}}, \bibinfo {author}
  {\bibfnamefont {K.}~\bibnamefont {Cunha}}, \bibinfo {author} {\bibfnamefont
  {J.~G.}\ \bibnamefont {Fernández-Trincado}}, \bibinfo {author}
  {\bibfnamefont {D.}~\bibnamefont {Horta}}, \bibinfo {author} {\bibfnamefont
  {R.~R.}\ \bibnamefont {Lane}}, \bibinfo {author} {\bibfnamefont {K.~L.}\
  \bibnamefont {Masters}}, \bibinfo {author} {\bibfnamefont {C.}~\bibnamefont
  {Nitschelm}},\ and\ \bibinfo {author} {\bibfnamefont {A.}~\bibnamefont
  {Roman-Lopes}},\ }\href {https://doi.org/10.1093/mnras/stac1151} {\bibfield
  {journal} {\bibinfo  {journal} {Monthly Notices of the Royal Astronomical
  Society}\ }\textbf {\bibinfo {volume} {513}},\ \bibinfo {pages} {4130}
  (\bibinfo {year} {2022})}\BibitemShut {NoStop}%
\bibitem [{\citenamefont {{Cantat-Gaudin}}\ \emph {et~al.}(2024)\citenamefont
  {{Cantat-Gaudin}}, \citenamefont {{Fouesneau}}, \citenamefont {{Rix}},
  \citenamefont {{Brown}}, \citenamefont {{Drimmel}}, \citenamefont
  {{Castro-Ginard}}, \citenamefont {{Khanna}}, \citenamefont {{Belokurov}},\
  and\ \citenamefont {{Casey}}}]{Cantat_2024}%
  \BibitemOpen
  \bibfield  {author} {\bibinfo {author} {\bibfnamefont {T.}~\bibnamefont
  {{Cantat-Gaudin}}}, \bibinfo {author} {\bibfnamefont {M.}~\bibnamefont
  {{Fouesneau}}}, \bibinfo {author} {\bibfnamefont {H.-W.}\ \bibnamefont
  {{Rix}}}, \bibinfo {author} {\bibfnamefont {A.~G.~A.}\ \bibnamefont
  {{Brown}}}, \bibinfo {author} {\bibfnamefont {R.}~\bibnamefont {{Drimmel}}},
  \bibinfo {author} {\bibfnamefont {A.}~\bibnamefont {{Castro-Ginard}}},
  \bibinfo {author} {\bibfnamefont {S.}~\bibnamefont {{Khanna}}}, \bibinfo
  {author} {\bibfnamefont {V.}~\bibnamefont {{Belokurov}}},\ and\ \bibinfo
  {author} {\bibfnamefont {A.~R.}\ \bibnamefont {{Casey}}},\ }\href
  {https://doi.org/10.1051/0004-6361/202348018} {\bibfield  {journal} {\bibinfo
   {journal} {Astronomy \& Astrophysics}\ }\textbf {\bibinfo {volume} {683}},\
  \bibinfo {eid} {A128} (\bibinfo {year} {2024})},\ \Eprint
  {https://arxiv.org/abs/2401.05023} {arXiv:2401.05023 [astro-ph.GA]}
  \BibitemShut {NoStop}%
\bibitem [{\citenamefont {Imig}\ \emph {et~al.}(2025)\citenamefont {Imig},
  \citenamefont {Holtzman}, \citenamefont {Zasowski}, \citenamefont {Lian},
  \citenamefont {Boardman}, \citenamefont {Stone-Martinez}, \citenamefont
  {Mackereth}, \citenamefont {Prescott}, \citenamefont {Beaton}, \citenamefont
  {Beers}, \citenamefont {Bizyaev}, \citenamefont {Blanton}, \citenamefont
  {Cunha}, \citenamefont {Fernández-Trincado}, \citenamefont {Fielder},
  \citenamefont {Hasselquist}, \citenamefont {Hayes}, \citenamefont {Haywood},
  \citenamefont {Jönsson}, \citenamefont {Lane}, \citenamefont {Majewski},
  \citenamefont {Mészáros}, \citenamefont {Minchev}, \citenamefont {Nidever},
  \citenamefont {Nitschelm},\ and\ \citenamefont {Sobeck}}]{Imig_2025}%
  \BibitemOpen
  \bibfield  {author} {\bibinfo {author} {\bibfnamefont {J.}~\bibnamefont
  {Imig}}, \bibinfo {author} {\bibfnamefont {J.~A.}\ \bibnamefont {Holtzman}},
  \bibinfo {author} {\bibfnamefont {G.}~\bibnamefont {Zasowski}}, \bibinfo
  {author} {\bibfnamefont {J.}~\bibnamefont {Lian}}, \bibinfo {author}
  {\bibfnamefont {N.~F.}\ \bibnamefont {Boardman}}, \bibinfo {author}
  {\bibfnamefont {A.}~\bibnamefont {Stone-Martinez}}, \bibinfo {author}
  {\bibfnamefont {J.~T.}\ \bibnamefont {Mackereth}}, \bibinfo {author}
  {\bibfnamefont {M.~K.~M.}\ \bibnamefont {Prescott}}, \bibinfo {author}
  {\bibfnamefont {R.~L.}\ \bibnamefont {Beaton}}, \bibinfo {author}
  {\bibfnamefont {T.~C.}\ \bibnamefont {Beers}}, \bibinfo {author}
  {\bibfnamefont {D.}~\bibnamefont {Bizyaev}}, \bibinfo {author} {\bibfnamefont
  {M.~R.}\ \bibnamefont {Blanton}}, \bibinfo {author} {\bibfnamefont
  {K.}~\bibnamefont {Cunha}}, \bibinfo {author} {\bibfnamefont {J.~G.}\
  \bibnamefont {Fernández-Trincado}}, \bibinfo {author} {\bibfnamefont
  {C.~E.}\ \bibnamefont {Fielder}}, \bibinfo {author} {\bibfnamefont
  {S.}~\bibnamefont {Hasselquist}}, \bibinfo {author} {\bibfnamefont {C.~R.}\
  \bibnamefont {Hayes}}, \bibinfo {author} {\bibfnamefont {M.}~\bibnamefont
  {Haywood}}, \bibinfo {author} {\bibfnamefont {H.}~\bibnamefont {Jönsson}},
  \bibinfo {author} {\bibfnamefont {R.~R.}\ \bibnamefont {Lane}}, \bibinfo
  {author} {\bibfnamefont {S.~R.}\ \bibnamefont {Majewski}}, \bibinfo {author}
  {\bibfnamefont {S.}~\bibnamefont {Mészáros}}, \bibinfo {author}
  {\bibfnamefont {I.}~\bibnamefont {Minchev}}, \bibinfo {author} {\bibfnamefont
  {D.~L.}\ \bibnamefont {Nidever}}, \bibinfo {author} {\bibfnamefont
  {C.}~\bibnamefont {Nitschelm}},\ and\ \bibinfo {author} {\bibfnamefont
  {J.}~\bibnamefont {Sobeck}},\ }\href
  {https://doi.org/10.3847/1538-4357/adf723} {\bibfield  {journal} {\bibinfo
  {journal} {The Astrophysical Journal}\ }\textbf {\bibinfo {volume} {990}},\
  \bibinfo {pages} {203} (\bibinfo {year} {2025})}\BibitemShut {NoStop}%
\bibitem [{\citenamefont {Yu}\ \emph {et~al.}(2021)\citenamefont {Yu},
  \citenamefont {Li}, \citenamefont {Chen}, \citenamefont {Huang},
  \citenamefont {Jia}, \citenamefont {Xiang}, \citenamefont {Yuan},
  \citenamefont {Shi}, \citenamefont {Wang},\ and\ \citenamefont
  {Liu}}]{Yu_2021}%
  \BibitemOpen
  \bibfield  {author} {\bibinfo {author} {\bibfnamefont {Z.}~\bibnamefont
  {Yu}}, \bibinfo {author} {\bibfnamefont {J.}~\bibnamefont {Li}}, \bibinfo
  {author} {\bibfnamefont {B.}~\bibnamefont {Chen}}, \bibinfo {author}
  {\bibfnamefont {Y.}~\bibnamefont {Huang}}, \bibinfo {author} {\bibfnamefont
  {S.}~\bibnamefont {Jia}}, \bibinfo {author} {\bibfnamefont {M.}~\bibnamefont
  {Xiang}}, \bibinfo {author} {\bibfnamefont {H.}~\bibnamefont {Yuan}},
  \bibinfo {author} {\bibfnamefont {J.}~\bibnamefont {Shi}}, \bibinfo {author}
  {\bibfnamefont {C.}~\bibnamefont {Wang}},\ and\ \bibinfo {author}
  {\bibfnamefont {X.}~\bibnamefont {Liu}},\ }\href
  {https://doi.org/10.3847/1538-4357/abf098} {\bibfield  {journal} {\bibinfo
  {journal} {The Astrophysical Journal}\ }\textbf {\bibinfo {volume} {912}},\
  \bibinfo {pages} {106} (\bibinfo {year} {2021})}\BibitemShut {NoStop}%
\bibitem [{\citenamefont {{Yu}}\ \emph {et~al.}(2025)\citenamefont {{Yu}},
  \citenamefont {{Chen}}, \citenamefont {{Lian}}, \citenamefont {{Wang}},\ and\
  \citenamefont {{Liu}}}]{Yu_2025}%
  \BibitemOpen
  \bibfield  {author} {\bibinfo {author} {\bibfnamefont {Z.}~\bibnamefont
  {{Yu}}}, \bibinfo {author} {\bibfnamefont {B.}~\bibnamefont {{Chen}}},
  \bibinfo {author} {\bibfnamefont {J.}~\bibnamefont {{Lian}}}, \bibinfo
  {author} {\bibfnamefont {C.}~\bibnamefont {{Wang}}},\ and\ \bibinfo {author}
  {\bibfnamefont {X.}~\bibnamefont {{Liu}}},\ }\href
  {https://doi.org/10.3847/1538-3881/ad9582} {\bibfield  {journal} {\bibinfo
  {journal} {The Astronomical Journal}\ }\textbf {\bibinfo {volume} {169}},\
  \bibinfo {eid} {61} (\bibinfo {year} {2025})},\ \Eprint
  {https://arxiv.org/abs/2412.14743} {arXiv:2412.14743 [astro-ph.GA]}
  \BibitemShut {NoStop}%
\bibitem [{\citenamefont {{Lian}}\ \emph {et~al.}(2024)\citenamefont {{Lian}},
  \citenamefont {{Zasowski}}, \citenamefont {{Chen}}, \citenamefont {{Imig}},
  \citenamefont {{Wang}}, \citenamefont {{Boardman}},\ and\ \citenamefont
  {{Liu}}}]{Lian_2024}%
  \BibitemOpen
  \bibfield  {author} {\bibinfo {author} {\bibfnamefont {J.}~\bibnamefont
  {{Lian}}}, \bibinfo {author} {\bibfnamefont {G.}~\bibnamefont {{Zasowski}}},
  \bibinfo {author} {\bibfnamefont {B.}~\bibnamefont {{Chen}}}, \bibinfo
  {author} {\bibfnamefont {J.}~\bibnamefont {{Imig}}}, \bibinfo {author}
  {\bibfnamefont {T.}~\bibnamefont {{Wang}}}, \bibinfo {author} {\bibfnamefont
  {N.}~\bibnamefont {{Boardman}}},\ and\ \bibinfo {author} {\bibfnamefont
  {X.}~\bibnamefont {{Liu}}},\ }\href
  {https://doi.org/10.1038/s41550-024-02315-7} {\bibfield  {journal} {\bibinfo
  {journal} {Nature Astronomy}\ }\textbf {\bibinfo {volume} {8}},\ \bibinfo
  {pages} {1302} (\bibinfo {year} {2024})},\ \Eprint
  {https://arxiv.org/abs/2406.05604} {arXiv:2406.05604 [astro-ph.GA]}
  \BibitemShut {NoStop}%
\bibitem [{\citenamefont {{Lian}}\ \emph {et~al.}(2025)\citenamefont {{Lian}},
  \citenamefont {{Wang}}, \citenamefont {{Feng}}, \citenamefont {{Huang}},\
  and\ \citenamefont {{Guo}}}]{Lian_2025}%
  \BibitemOpen
  \bibfield  {author} {\bibinfo {author} {\bibfnamefont {J.}~\bibnamefont
  {{Lian}}}, \bibinfo {author} {\bibfnamefont {T.}~\bibnamefont {{Wang}}},
  \bibinfo {author} {\bibfnamefont {Q.}~\bibnamefont {{Feng}}}, \bibinfo
  {author} {\bibfnamefont {Y.}~\bibnamefont {{Huang}}},\ and\ \bibinfo {author}
  {\bibfnamefont {H.}~\bibnamefont {{Guo}}},\ }\href
  {https://doi.org/10.3847/2041-8213/adfc73} {\bibfield  {journal} {\bibinfo
  {journal} {The Astrophysical Journal Letter}\ }\textbf {\bibinfo {volume}
  {990}},\ \bibinfo {eid} {L37} (\bibinfo {year} {2025})},\ \Eprint
  {https://arxiv.org/abs/2508.13665} {arXiv:2508.13665 [astro-ph.GA]}
  \BibitemShut {NoStop}%
\bibitem [{\citenamefont {{Ou}}\ \emph {et~al.}(2024)\citenamefont {{Ou}},
  \citenamefont {{Eilers}}, \citenamefont {{Necib}},\ and\ \citenamefont
  {{Frebel}}}]{Ou_2024}%
  \BibitemOpen
  \bibfield  {author} {\bibinfo {author} {\bibfnamefont {X.}~\bibnamefont
  {{Ou}}}, \bibinfo {author} {\bibfnamefont {A.-C.}\ \bibnamefont {{Eilers}}},
  \bibinfo {author} {\bibfnamefont {L.}~\bibnamefont {{Necib}}},\ and\ \bibinfo
  {author} {\bibfnamefont {A.}~\bibnamefont {{Frebel}}},\ }\href
  {https://doi.org/10.1093/mnras/stae034} {\bibfield  {journal} {\bibinfo
  {journal} {Monthly Notices of the Royal Astronomical Society}\ }\textbf
  {\bibinfo {volume} {528}},\ \bibinfo {pages} {693} (\bibinfo {year}
  {2024})},\ \Eprint {https://arxiv.org/abs/2303.12838} {arXiv:2303.12838
  [astro-ph.GA]} \BibitemShut {NoStop}%
\bibitem [{\citenamefont {Majewski}\ \emph {et~al.}(2017)\citenamefont
  {Majewski} \emph {et~al.}}]{Majewski_2017}%
  \BibitemOpen
  \bibfield  {author} {\bibinfo {author} {\bibfnamefont {S.~R.}\ \bibnamefont
  {Majewski}} \emph {et~al.},\ }\href
  {https://doi.org/10.3847/1538-3881/aa784d} {\bibfield  {journal} {\bibinfo
  {journal} {Astron. J.}\ }\textbf {\bibinfo {volume} {154}},\ \bibinfo {pages}
  {94} (\bibinfo {year} {2017})}\BibitemShut {NoStop}%
\bibitem [{\citenamefont {Vallenari}\ \emph {et~al.}(2023)\citenamefont
  {Vallenari} \emph {et~al.}}]{Gaia_2023}%
  \BibitemOpen
  \bibfield  {author} {\bibinfo {author} {\bibfnamefont {A.}~\bibnamefont
  {Vallenari}} \emph {et~al.},\ }\href
  {https://doi.org/10.1051/0004-6361/202243940} {\bibfield  {journal} {\bibinfo
   {journal} {Astron. Astrophys.}\ }\textbf {\bibinfo {volume} {674}},\
  \bibinfo {pages} {A1} (\bibinfo {year} {2023})}\BibitemShut {NoStop}%
\bibitem [{\citenamefont {Skrutskie}\ \emph {et~al.}(2006)\citenamefont
  {Skrutskie} \emph {et~al.}}]{2MASS}%
  \BibitemOpen
  \bibfield  {author} {\bibinfo {author} {\bibfnamefont {M.~F.}\ \bibnamefont
  {Skrutskie}} \emph {et~al.},\ }\href {https://doi.org/10.1086/498708}
  {\bibfield  {journal} {\bibinfo  {journal} {Astron. J.}\ }\textbf {\bibinfo
  {volume} {131}},\ \bibinfo {pages} {1163} (\bibinfo {year}
  {2006})}\BibitemShut {NoStop}%
\bibitem [{\citenamefont {Wright}\ \emph {et~al.}(2010)\citenamefont {Wright}
  \emph {et~al.}}]{Wright_2010}%
  \BibitemOpen
  \bibfield  {author} {\bibinfo {author} {\bibfnamefont {E.~L.}\ \bibnamefont
  {Wright}} \emph {et~al.},\ }\href
  {https://doi.org/10.1088/0004-6256/140/6/1868} {\bibfield  {journal}
  {\bibinfo  {journal} {Astron. J.}\ }\textbf {\bibinfo {volume} {140}},\
  \bibinfo {pages} {1868} (\bibinfo {year} {2010})}\BibitemShut {NoStop}%
\bibitem [{\citenamefont {{Eilers}}\ \emph {et~al.}(2019)\citenamefont
  {{Eilers}}, \citenamefont {{Hogg}}, \citenamefont {{Rix}},\ and\
  \citenamefont {{Ness}}}]{Eilers_2019}%
  \BibitemOpen
  \bibfield  {author} {\bibinfo {author} {\bibfnamefont {A.-C.}\ \bibnamefont
  {{Eilers}}}, \bibinfo {author} {\bibfnamefont {D.~W.}\ \bibnamefont
  {{Hogg}}}, \bibinfo {author} {\bibfnamefont {H.-W.}\ \bibnamefont {{Rix}}},\
  and\ \bibinfo {author} {\bibfnamefont {M.~K.}\ \bibnamefont {{Ness}}},\
  }\href {https://doi.org/10.3847/1538-4357/aaf648} {\bibfield  {journal}
  {\bibinfo  {journal} {The Astrophysical Journal}\ }\textbf {\bibinfo {volume}
  {871}},\ \bibinfo {eid} {120} (\bibinfo {year} {2019})},\ \Eprint
  {https://arxiv.org/abs/1810.09466} {arXiv:1810.09466 [astro-ph.GA]}
  \BibitemShut {NoStop}%
\bibitem [{\citenamefont {Zhou}\ \emph {et~al.}(2023)\citenamefont {Zhou},
  \citenamefont {Li}, \citenamefont {Huang},\ and\ \citenamefont
  {Zhang}}]{Zhou_2023}%
  \BibitemOpen
  \bibfield  {author} {\bibinfo {author} {\bibfnamefont {Y.}~\bibnamefont
  {Zhou}}, \bibinfo {author} {\bibfnamefont {X.}~\bibnamefont {Li}}, \bibinfo
  {author} {\bibfnamefont {Y.}~\bibnamefont {Huang}},\ and\ \bibinfo {author}
  {\bibfnamefont {H.}~\bibnamefont {Zhang}},\ }\href
  {https://doi.org/10.3847/1538-4357/acadd9} {\bibfield  {journal} {\bibinfo
  {journal} {The Astrophysical Journal}\ }\textbf {\bibinfo {volume} {946}},\
  \bibinfo {pages} {73} (\bibinfo {year} {2023})}\BibitemShut {NoStop}%
\bibitem [{\citenamefont {{Buch}}\ \emph {et~al.}(2019)\citenamefont {{Buch}},
  \citenamefont {{Leung}},\ and\ \citenamefont {{Fan}}}]{Buch_2019}%
  \BibitemOpen
  \bibfield  {author} {\bibinfo {author} {\bibfnamefont {J.}~\bibnamefont
  {{Buch}}}, \bibinfo {author} {\bibfnamefont {J.~S.~C.}\ \bibnamefont
  {{Leung}}},\ and\ \bibinfo {author} {\bibfnamefont {J.}~\bibnamefont
  {{Fan}}},\ }\href {https://doi.org/10.1088/1475-7516/2019/04/026} {\bibfield
  {journal} {\bibinfo  {journal} {Journal of Cosmology and Astroparticle
  Physics}\ }\textbf {\bibinfo {volume} {2019}}\bibfield  {number} {\bibinfo
  {number} { (4)},\ \bibinfo {eid} {026}},\ }\Eprint
  {https://arxiv.org/abs/1808.05603} {arXiv:1808.05603 [astro-ph.GA]}
  \BibitemShut {NoStop}%
\bibitem [{\citenamefont {{Cheng}}\ \emph {et~al.}(2024)\citenamefont
  {{Cheng}}, \citenamefont {{Anguiano}}, \citenamefont {{Majewski}},\ and\
  \citenamefont {{Arras}}}]{Cheng_2024}%
  \BibitemOpen
  \bibfield  {author} {\bibinfo {author} {\bibfnamefont {X.}~\bibnamefont
  {{Cheng}}}, \bibinfo {author} {\bibfnamefont {B.}~\bibnamefont {{Anguiano}}},
  \bibinfo {author} {\bibfnamefont {S.~R.}\ \bibnamefont {{Majewski}}},\ and\
  \bibinfo {author} {\bibfnamefont {P.}~\bibnamefont {{Arras}}},\ }\href
  {https://doi.org/10.1093/mnras/stad3013} {\bibfield  {journal} {\bibinfo
  {journal} {Monthly Notices of the Royal Astronomical Society}\ }\textbf
  {\bibinfo {volume} {527}},\ \bibinfo {pages} {959} (\bibinfo {year}
  {2024})},\ \Eprint {https://arxiv.org/abs/2309.17405} {arXiv:2309.17405
  [astro-ph.GA]} \BibitemShut {NoStop}%
\bibitem [{\citenamefont {{S{\"o}ding}}\ \emph {et~al.}(2025)\citenamefont
  {{S{\"o}ding}}, \citenamefont {{Bartel}},\ and\ \citenamefont
  {{Mertsch}}}]{Soding_2025}%
  \BibitemOpen
  \bibfield  {author} {\bibinfo {author} {\bibfnamefont {L.}~\bibnamefont
  {{S{\"o}ding}}}, \bibinfo {author} {\bibfnamefont {R.~L.}\ \bibnamefont
  {{Bartel}}},\ and\ \bibinfo {author} {\bibfnamefont {P.}~\bibnamefont
  {{Mertsch}}},\ }\href {https://doi.org/10.1093/mnras/staf1391} {\bibfield
  {journal} {\bibinfo  {journal} {Monthly Notices of the Royal Astronomical
  Society}\ }\textbf {\bibinfo {volume} {542}},\ \bibinfo {pages} {2987}
  (\bibinfo {year} {2025})},\ \Eprint {https://arxiv.org/abs/2506.02956}
  {arXiv:2506.02956 [astro-ph.GA]} \BibitemShut {NoStop}%
\bibitem [{\citenamefont {{Ou}}\ \emph {et~al.}(2025)\citenamefont {{Ou}},
  \citenamefont {{Necib}}, \citenamefont {{Wetzel}}, \citenamefont {{Frebel}},
  \citenamefont {{Bailin}},\ and\ \citenamefont {{Oeur}}}]{Ou_2025}%
  \BibitemOpen
  \bibfield  {author} {\bibinfo {author} {\bibfnamefont {X.}~\bibnamefont
  {{Ou}}}, \bibinfo {author} {\bibfnamefont {L.}~\bibnamefont {{Necib}}},
  \bibinfo {author} {\bibfnamefont {A.}~\bibnamefont {{Wetzel}}}, \bibinfo
  {author} {\bibfnamefont {A.}~\bibnamefont {{Frebel}}}, \bibinfo {author}
  {\bibfnamefont {J.}~\bibnamefont {{Bailin}}},\ and\ \bibinfo {author}
  {\bibfnamefont {M.}~\bibnamefont {{Oeur}}},\ }\href
  {https://doi.org/10.3847/1538-4357/ae0b63} {\bibfield  {journal} {\bibinfo
  {journal} {The Astrophysical Journal}\ }\textbf {\bibinfo {volume} {994}},\
  \bibinfo {eid} {128} (\bibinfo {year} {2025})},\ \Eprint
  {https://arxiv.org/abs/2503.05877} {arXiv:2503.05877 [astro-ph.GA]}
  \BibitemShut {NoStop}%
\bibitem [{\citenamefont {McMillan}(2016)}]{McMillan_2016}%
  \BibitemOpen
  \bibfield  {author} {\bibinfo {author} {\bibfnamefont {P.~J.}\ \bibnamefont
  {McMillan}},\ }\href {https://doi.org/10.1093/mnras/stw2759} {\bibfield
  {journal} {\bibinfo  {journal} {Monthly Notices of the Royal Astronomical
  Society}\ }\textbf {\bibinfo {volume} {465}},\ \bibinfo {pages} {76}
  (\bibinfo {year} {2016})}\BibitemShut {NoStop}%
\bibitem [{\citenamefont {Sormani}\ \emph {et~al.}(2022)\citenamefont
  {Sormani}, \citenamefont {Gerhard}, \citenamefont {Portail}, \citenamefont
  {Vasiliev},\ and\ \citenamefont {Clarke}}]{Sormani_2022}%
  \BibitemOpen
  \bibfield  {author} {\bibinfo {author} {\bibfnamefont {M.~C.}\ \bibnamefont
  {Sormani}}, \bibinfo {author} {\bibfnamefont {O.}~\bibnamefont {Gerhard}},
  \bibinfo {author} {\bibfnamefont {M.}~\bibnamefont {Portail}}, \bibinfo
  {author} {\bibfnamefont {E.}~\bibnamefont {Vasiliev}},\ and\ \bibinfo
  {author} {\bibfnamefont {J.}~\bibnamefont {Clarke}},\ }\href
  {https://doi.org/10.1093/mnrasl/slac046} {\bibfield  {journal} {\bibinfo
  {journal} {Monthly Notices of the Royal Astronomical Society: Letters}\
  }\textbf {\bibinfo {volume} {514}},\ \bibinfo {pages} {L1} (\bibinfo {year}
  {2022})}\BibitemShut {NoStop}%
\bibitem [{\citenamefont {{Zhao}}(1996)}]{Zhao_1996}%
  \BibitemOpen
  \bibfield  {author} {\bibinfo {author} {\bibfnamefont {H.}~\bibnamefont
  {{Zhao}}},\ }\href {https://doi.org/10.1093/mnras/278.2.488} {\bibfield
  {journal} {\bibinfo  {journal} {Monthly Notices of the Royal Astronomical
  Society}\ }\textbf {\bibinfo {volume} {278}},\ \bibinfo {pages} {488}
  (\bibinfo {year} {1996})},\ \Eprint {https://arxiv.org/abs/astro-ph/9509122}
  {arXiv:astro-ph/9509122 [astro-ph]} \BibitemShut {NoStop}%
\bibitem [{\citenamefont {{Navarro}}\ \emph {et~al.}(1997)\citenamefont
  {{Navarro}}, \citenamefont {{Frenk}},\ and\ \citenamefont
  {{White}}}]{Navarro_1997}%
  \BibitemOpen
  \bibfield  {author} {\bibinfo {author} {\bibfnamefont {J.~F.}\ \bibnamefont
  {{Navarro}}}, \bibinfo {author} {\bibfnamefont {C.~S.}\ \bibnamefont
  {{Frenk}}},\ and\ \bibinfo {author} {\bibfnamefont {S.~D.~M.}\ \bibnamefont
  {{White}}},\ }\href {https://doi.org/10.1086/304888} {\bibfield  {journal}
  {\bibinfo  {journal} {The Astrophysical Journal}\ }\textbf {\bibinfo {volume}
  {490}},\ \bibinfo {pages} {493} (\bibinfo {year} {1997})},\ \Eprint
  {https://arxiv.org/abs/astro-ph/9611107} {arXiv:astro-ph/9611107 [astro-ph]}
  \BibitemShut {NoStop}%
\bibitem [{\citenamefont {{Einasto}}(1965)}]{Einasto_1965}%
  \BibitemOpen
  \bibfield  {author} {\bibinfo {author} {\bibfnamefont {J.}~\bibnamefont
  {{Einasto}}},\ }\href@noop {} {\bibfield  {journal} {\bibinfo  {journal}
  {Trudy Astrofizicheskogo Instituta Alma-Ata}\ }\textbf {\bibinfo {volume}
  {5}},\ \bibinfo {pages} {87} (\bibinfo {year} {1965})}\BibitemShut {NoStop}%
\bibitem [{\citenamefont {{Moffat}}(2006)}]{Moffat_2006}%
  \BibitemOpen
  \bibfield  {author} {\bibinfo {author} {\bibfnamefont {J.~W.}\ \bibnamefont
  {{Moffat}}},\ }\href {https://doi.org/10.1088/1475-7516/2006/03/004}
  {\bibfield  {journal} {\bibinfo  {journal} {Journal of Cosmology and
  Astroparticle Physics}\ }\textbf {\bibinfo {volume} {2006}}\bibfield
  {number} {\bibinfo  {number} { (3)},\ \bibinfo {eid} {004}},\ }\Eprint
  {https://arxiv.org/abs/gr-qc/0506021} {arXiv:gr-qc/0506021 [gr-qc]}
  \BibitemShut {NoStop}%
\bibitem [{\citenamefont {Dutton}\ and\ \citenamefont
  {Macci{\`o}}(2014)}]{Dutton_2014}%
  \BibitemOpen
  \bibfield  {author} {\bibinfo {author} {\bibfnamefont {A.~A.}\ \bibnamefont
  {Dutton}}\ and\ \bibinfo {author} {\bibfnamefont {A.~V.}\ \bibnamefont
  {Macci{\`o}}},\ }\href@noop {} {\bibfield  {journal} {\bibinfo  {journal}
  {Monthly Notices of the Royal Astronomical Society}\ }\textbf {\bibinfo
  {volume} {441}},\ \bibinfo {pages} {3359} (\bibinfo {year}
  {2014})}\BibitemShut {NoStop}%
\bibitem [{\citenamefont {Macci{\`o}}\ \emph {et~al.}(2008)\citenamefont
  {Macci{\`o}}, \citenamefont {Dutton},\ and\ \citenamefont {van~den
  Bosch}}]{Maccio_2008}%
  \BibitemOpen
  \bibfield  {author} {\bibinfo {author} {\bibfnamefont {A.~V.}\ \bibnamefont
  {Macci{\`o}}}, \bibinfo {author} {\bibfnamefont {A.~A.}\ \bibnamefont
  {Dutton}},\ and\ \bibinfo {author} {\bibfnamefont {F.~C.}\ \bibnamefont
  {van~den Bosch}},\ }\href@noop {} {\bibfield  {journal} {\bibinfo  {journal}
  {Monthly Notices of the Royal Astronomical Society}\ }\textbf {\bibinfo
  {volume} {391}},\ \bibinfo {pages} {1940} (\bibinfo {year}
  {2008})}\BibitemShut {NoStop}%
\bibitem [{\citenamefont {Di~Cintio}\ \emph {et~al.}(2014)\citenamefont
  {Di~Cintio}, \citenamefont {Brook}, \citenamefont {Dutton}, \citenamefont
  {Macci{\`o}}, \citenamefont {Stinson},\ and\ \citenamefont
  {Knebe}}]{DiCintio_2014}%
  \BibitemOpen
  \bibfield  {author} {\bibinfo {author} {\bibfnamefont {A.}~\bibnamefont
  {Di~Cintio}}, \bibinfo {author} {\bibfnamefont {C.~B.}\ \bibnamefont
  {Brook}}, \bibinfo {author} {\bibfnamefont {A.~A.}\ \bibnamefont {Dutton}},
  \bibinfo {author} {\bibfnamefont {A.~V.}\ \bibnamefont {Macci{\`o}}},
  \bibinfo {author} {\bibfnamefont {G.~S.}\ \bibnamefont {Stinson}},\ and\
  \bibinfo {author} {\bibfnamefont {A.}~\bibnamefont {Knebe}},\ }\href@noop {}
  {\bibfield  {journal} {\bibinfo  {journal} {Monthly Notices of the Royal
  Astronomical Society}\ }\textbf {\bibinfo {volume} {441}},\ \bibinfo {pages}
  {2986} (\bibinfo {year} {2014})}\BibitemShut {NoStop}%
\bibitem [{\citenamefont {Freundlich}\ \emph {et~al.}(2020)\citenamefont
  {Freundlich}, \citenamefont {Dekel}, \citenamefont {Jiang}, \citenamefont
  {Ishai}, \citenamefont {Cornuault}, \citenamefont {Lapiner}, \citenamefont
  {Dutton},\ and\ \citenamefont {Macci{\`o}}}]{Freundlich_2020}%
  \BibitemOpen
  \bibfield  {author} {\bibinfo {author} {\bibfnamefont {J.}~\bibnamefont
  {Freundlich}}, \bibinfo {author} {\bibfnamefont {A.}~\bibnamefont {Dekel}},
  \bibinfo {author} {\bibfnamefont {F.}~\bibnamefont {Jiang}}, \bibinfo
  {author} {\bibfnamefont {G.}~\bibnamefont {Ishai}}, \bibinfo {author}
  {\bibfnamefont {N.}~\bibnamefont {Cornuault}}, \bibinfo {author}
  {\bibfnamefont {S.}~\bibnamefont {Lapiner}}, \bibinfo {author} {\bibfnamefont
  {A.~A.}\ \bibnamefont {Dutton}},\ and\ \bibinfo {author} {\bibfnamefont
  {A.~V.}\ \bibnamefont {Macci{\`o}}},\ }\href@noop {} {\bibfield  {journal}
  {\bibinfo  {journal} {Monthly Notices of the Royal Astronomical Society}\
  }\textbf {\bibinfo {volume} {491}},\ \bibinfo {pages} {1190} (\bibinfo {year}
  {2020})}\BibitemShut {NoStop}%
\bibitem [{\citenamefont {Milgrom}\ and\ \citenamefont
  {Sanders}(2008)}]{Milgrom_2008}%
  \BibitemOpen
  \bibfield  {author} {\bibinfo {author} {\bibfnamefont {M.}~\bibnamefont
  {Milgrom}}\ and\ \bibinfo {author} {\bibfnamefont {R.~H.}\ \bibnamefont
  {Sanders}},\ }\href {https://doi.org/10.1086/529119} {\bibfield  {journal}
  {\bibinfo  {journal} {The Astrophysical Journal}\ }\textbf {\bibinfo {volume}
  {678}},\ \bibinfo {pages} {131–143} (\bibinfo {year} {2008})}\BibitemShut
  {NoStop}%
\bibitem [{\citenamefont {{Adibekyan}}\ \emph {et~al.}(2012)\citenamefont
  {{Adibekyan}}, \citenamefont {{Sousa}}, \citenamefont {{Santos}},
  \citenamefont {{Delgado Mena}}, \citenamefont {{Gonz{\'a}lez Hern{\'a}ndez}},
  \citenamefont {{Israelian}}, \citenamefont {{Mayor}},\ and\ \citenamefont
  {{Khachatryan}}}]{Adibekyan_2012}%
  \BibitemOpen
  \bibfield  {author} {\bibinfo {author} {\bibfnamefont {V.~Z.}\ \bibnamefont
  {{Adibekyan}}}, \bibinfo {author} {\bibfnamefont {S.~G.}\ \bibnamefont
  {{Sousa}}}, \bibinfo {author} {\bibfnamefont {N.~C.}\ \bibnamefont
  {{Santos}}}, \bibinfo {author} {\bibfnamefont {E.}~\bibnamefont {{Delgado
  Mena}}}, \bibinfo {author} {\bibfnamefont {J.~I.}\ \bibnamefont
  {{Gonz{\'a}lez Hern{\'a}ndez}}}, \bibinfo {author} {\bibfnamefont
  {G.}~\bibnamefont {{Israelian}}}, \bibinfo {author} {\bibfnamefont
  {M.}~\bibnamefont {{Mayor}}},\ and\ \bibinfo {author} {\bibfnamefont
  {G.}~\bibnamefont {{Khachatryan}}},\ }\href
  {https://doi.org/10.1051/0004-6361/201219401} {\bibfield  {journal} {\bibinfo
   {journal} {Astronomy \& Astrophysics}\ }\textbf {\bibinfo {volume} {545}},\
  \bibinfo {eid} {A32} (\bibinfo {year} {2012})},\ \Eprint
  {https://arxiv.org/abs/1207.2388} {arXiv:1207.2388 [astro-ph.EP]}
  \BibitemShut {NoStop}%
\bibitem [{\citenamefont {Binney}\ and\ \citenamefont
  {Tremaine}(2008)}]{Binney_2008}%
  \BibitemOpen
  \bibfield  {author} {\bibinfo {author} {\bibfnamefont {J.}~\bibnamefont
  {Binney}}\ and\ \bibinfo {author} {\bibfnamefont {S.}~\bibnamefont
  {Tremaine}},\ }\href@noop {} {\emph {\bibinfo {title} {Galactic Dynamics}}},\
  \bibinfo {edition} {2nd}\ ed.\ (\bibinfo  {publisher} {Princeton University
  Press},\ \bibinfo {year} {2008})\ \bibinfo {note} {princeton Series in
  Astrophysics}\BibitemShut {NoStop}%
\bibitem [{\citenamefont {{Aigrain}}\ and\ \citenamefont
  {{Foreman-Mackey}}(2023)}]{Aigrain_2023}%
  \BibitemOpen
  \bibfield  {author} {\bibinfo {author} {\bibfnamefont {S.}~\bibnamefont
  {{Aigrain}}}\ and\ \bibinfo {author} {\bibfnamefont {D.}~\bibnamefont
  {{Foreman-Mackey}}},\ }\href
  {https://doi.org/10.1146/annurev-astro-052920-103508} {\bibfield  {journal}
  {\bibinfo  {journal} {Annual Review of Astronomy and Astrophysics}\ }\textbf
  {\bibinfo {volume} {61}},\ \bibinfo {pages} {329} (\bibinfo {year} {2023})},\
  \Eprint {https://arxiv.org/abs/2209.08940} {arXiv:2209.08940 [astro-ph.IM]}
  \BibitemShut {NoStop}%
\bibitem [{\citenamefont {{Guo}}\ \emph {et~al.}(2020)\citenamefont {{Guo}},
  \citenamefont {{Liu}}, \citenamefont {{Mao}}, \citenamefont {{Xue}},
  \citenamefont {{Long}},\ and\ \citenamefont {{Zhang}}}]{Guo_2020}%
  \BibitemOpen
  \bibfield  {author} {\bibinfo {author} {\bibfnamefont {R.}~\bibnamefont
  {{Guo}}}, \bibinfo {author} {\bibfnamefont {C.}~\bibnamefont {{Liu}}},
  \bibinfo {author} {\bibfnamefont {S.}~\bibnamefont {{Mao}}}, \bibinfo
  {author} {\bibfnamefont {X.-X.}\ \bibnamefont {{Xue}}}, \bibinfo {author}
  {\bibfnamefont {R.~J.}\ \bibnamefont {{Long}}},\ and\ \bibinfo {author}
  {\bibfnamefont {L.}~\bibnamefont {{Zhang}}},\ }\href
  {https://doi.org/10.1093/mnras/staa1483} {\bibfield  {journal} {\bibinfo
  {journal} {Monthly Notices of the Royal Astronomical Society}\ }\textbf
  {\bibinfo {volume} {495}},\ \bibinfo {pages} {4828} (\bibinfo {year}
  {2020})},\ \Eprint {https://arxiv.org/abs/2005.12018} {arXiv:2005.12018
  [astro-ph.GA]} \BibitemShut {NoStop}%
\bibitem [{\citenamefont {{Bailer-Jones}}\ \emph {et~al.}(2021)\citenamefont
  {{Bailer-Jones}}, \citenamefont {{Rybizki}}, \citenamefont {{Fouesneau}},
  \citenamefont {{Demleitner}},\ and\ \citenamefont {{Andrae}}}]{Bailer_2021}%
  \BibitemOpen
  \bibfield  {author} {\bibinfo {author} {\bibfnamefont {C.~A.~L.}\
  \bibnamefont {{Bailer-Jones}}}, \bibinfo {author} {\bibfnamefont
  {J.}~\bibnamefont {{Rybizki}}}, \bibinfo {author} {\bibfnamefont
  {M.}~\bibnamefont {{Fouesneau}}}, \bibinfo {author} {\bibfnamefont
  {M.}~\bibnamefont {{Demleitner}}},\ and\ \bibinfo {author} {\bibfnamefont
  {R.}~\bibnamefont {{Andrae}}},\ }\href
  {https://doi.org/10.3847/1538-3881/abd806} {\bibfield  {journal} {\bibinfo
  {journal} {The Astronomical Journal}\ }\textbf {\bibinfo {volume} {161}},\
  \bibinfo {eid} {147} (\bibinfo {year} {2021})},\ \Eprint
  {https://arxiv.org/abs/2012.05220} {arXiv:2012.05220 [astro-ph.SR]}
  \BibitemShut {NoStop}%
\bibitem [{\citenamefont {{Milgrom}}(2010)}]{Milgrom_2010}%
  \BibitemOpen
  \bibfield  {author} {\bibinfo {author} {\bibfnamefont {M.}~\bibnamefont
  {{Milgrom}}},\ }\href {https://doi.org/10.1111/j.1365-2966.2009.16184.x}
  {\bibfield  {journal} {\bibinfo  {journal} {Monthly Notices of the Royal
  Astronomical Society}\ }\textbf {\bibinfo {volume} {403}},\ \bibinfo {pages}
  {886} (\bibinfo {year} {2010})},\ \Eprint {https://arxiv.org/abs/0911.5464}
  {arXiv:0911.5464 [astro-ph.CO]} \BibitemShut {NoStop}%
\bibitem [{\citenamefont {Zhao}(2007)}]{Zhao_2007}%
  \BibitemOpen
  \bibfield  {author} {\bibinfo {author} {\bibfnamefont {H.}~\bibnamefont
  {Zhao}},\ }\href {https://doi.org/10.1086/524731} {\bibfield  {journal}
  {\bibinfo  {journal} {The Astrophysical Journal}\ }\textbf {\bibinfo {volume}
  {671}},\ \bibinfo {pages} {L1–L4} (\bibinfo {year} {2007})}\BibitemShut
  {NoStop}%
\bibitem [{\citenamefont {McKee}\ \emph {et~al.}(2015)\citenamefont {McKee},
  \citenamefont {Parravano},\ and\ \citenamefont {Hollenbach}}]{McKee_2015}%
  \BibitemOpen
  \bibfield  {author} {\bibinfo {author} {\bibfnamefont {C.~F.}\ \bibnamefont
  {McKee}}, \bibinfo {author} {\bibfnamefont {A.}~\bibnamefont {Parravano}},\
  and\ \bibinfo {author} {\bibfnamefont {D.~J.}\ \bibnamefont {Hollenbach}},\
  }\href {https://doi.org/10.1088/0004-637X/814/1/13} {\bibfield  {journal}
  {\bibinfo  {journal} {The Astrophysical Journal}\ }\textbf {\bibinfo {volume}
  {814}},\ \bibinfo {pages} {13} (\bibinfo {year} {2015})}\BibitemShut
  {NoStop}%
\bibitem [{\citenamefont {{Trotta}}(2008)}]{Trotta_2008}%
  \BibitemOpen
  \bibfield  {author} {\bibinfo {author} {\bibfnamefont {R.}~\bibnamefont
  {{Trotta}}},\ }\href {https://doi.org/10.1080/00107510802066753} {\bibfield
  {journal} {\bibinfo  {journal} {Contemporary Physics}\ }\textbf {\bibinfo
  {volume} {49}},\ \bibinfo {pages} {71} (\bibinfo {year} {2008})},\ \Eprint
  {https://arxiv.org/abs/0803.4089} {arXiv:0803.4089 [astro-ph]} \BibitemShut
  {NoStop}%
\bibitem [{\citenamefont {Burnham}\ and\ \citenamefont
  {Anderson}(2002)}]{Burnham_2002}%
  \BibitemOpen
  \bibfield  {author} {\bibinfo {author} {\bibfnamefont {K.~P.}\ \bibnamefont
  {Burnham}}\ and\ \bibinfo {author} {\bibfnamefont {D.~R.}\ \bibnamefont
  {Anderson}},\ }\href@noop {} {\emph {\bibinfo {title} {Model Selection and
  Multimodel Inference: A Practical Information-Theoretic Approach}}}\
  (\bibinfo  {publisher} {Springer-Verlag},\ \bibinfo {year}
  {2002})\BibitemShut {NoStop}%
\bibitem [{\citenamefont {Gelman}\ \emph {et~al.}(2013)\citenamefont {Gelman},
  \citenamefont {Carlin}, \citenamefont {Stern}, \citenamefont {Dunson},
  \citenamefont {Vehtari},\ and\ \citenamefont {Rubin}}]{Gelman_2013}%
  \BibitemOpen
  \bibfield  {author} {\bibinfo {author} {\bibfnamefont {A.}~\bibnamefont
  {Gelman}}, \bibinfo {author} {\bibfnamefont {J.~B.}\ \bibnamefont {Carlin}},
  \bibinfo {author} {\bibfnamefont {H.~S.}\ \bibnamefont {Stern}}, \bibinfo
  {author} {\bibfnamefont {D.~B.}\ \bibnamefont {Dunson}}, \bibinfo {author}
  {\bibfnamefont {A.}~\bibnamefont {Vehtari}},\ and\ \bibinfo {author}
  {\bibfnamefont {D.~B.}\ \bibnamefont {Rubin}},\ }\href@noop {} {\emph
  {\bibinfo {title} {Bayesian Data Analysis}}},\ \bibinfo {edition} {3rd}\ ed.\
  (\bibinfo  {publisher} {CRC Press},\ \bibinfo {year} {2013})\BibitemShut
  {NoStop}%
\bibitem [{\citenamefont {{Foreman-Mackey}}\ \emph {et~al.}(2013)\citenamefont
  {{Foreman-Mackey}}, \citenamefont {{Hogg}}, \citenamefont {{Lang}},\ and\
  \citenamefont {{Goodman}}}]{Foreman_2013}%
  \BibitemOpen
  \bibfield  {author} {\bibinfo {author} {\bibfnamefont {D.}~\bibnamefont
  {{Foreman-Mackey}}}, \bibinfo {author} {\bibfnamefont {D.~W.}\ \bibnamefont
  {{Hogg}}}, \bibinfo {author} {\bibfnamefont {D.}~\bibnamefont {{Lang}}},\
  and\ \bibinfo {author} {\bibfnamefont {J.}~\bibnamefont {{Goodman}}},\ }\href
  {https://doi.org/10.1086/670067} {\bibfield  {journal} {\bibinfo  {journal}
  {Publications of the Astronomical Society of the Pacific}\ }\textbf {\bibinfo
  {volume} {125}},\ \bibinfo {pages} {306} (\bibinfo {year} {2013})},\ \Eprint
  {https://arxiv.org/abs/1202.3665} {arXiv:1202.3665 [astro-ph.IM]}
  \BibitemShut {NoStop}%
\bibitem [{\citenamefont {{Schwarz}}(1978)}]{Schwarz_1978}%
  \BibitemOpen
  \bibfield  {author} {\bibinfo {author} {\bibfnamefont {G.}~\bibnamefont
  {{Schwarz}}},\ }\href@noop {} {\bibfield  {journal} {\bibinfo  {journal}
  {Annals of Statistics}\ }\textbf {\bibinfo {volume} {6}},\ \bibinfo {pages}
  {461} (\bibinfo {year} {1978})}\BibitemShut {NoStop}%
\bibitem [{\citenamefont {{Huang}}\ \emph {et~al.}(2016)\citenamefont
  {{Huang}}, \citenamefont {{Liu}}, \citenamefont {{Yuan}}, \citenamefont
  {{Xiang}}, \citenamefont {{Zhang}}, \citenamefont {{Chen}}, \citenamefont
  {{Ren}}, \citenamefont {{Wang}}, \citenamefont {{Zhang}}, \citenamefont
  {{Hou}}, \citenamefont {{Wang}},\ and\ \citenamefont {{Cao}}}]{Huang_2016}%
  \BibitemOpen
  \bibfield  {author} {\bibinfo {author} {\bibfnamefont {Y.}~\bibnamefont
  {{Huang}}}, \bibinfo {author} {\bibfnamefont {X.-W.}\ \bibnamefont {{Liu}}},
  \bibinfo {author} {\bibfnamefont {H.-B.}\ \bibnamefont {{Yuan}}}, \bibinfo
  {author} {\bibfnamefont {M.-S.}\ \bibnamefont {{Xiang}}}, \bibinfo {author}
  {\bibfnamefont {H.-W.}\ \bibnamefont {{Zhang}}}, \bibinfo {author}
  {\bibfnamefont {B.-Q.}\ \bibnamefont {{Chen}}}, \bibinfo {author}
  {\bibfnamefont {J.-J.}\ \bibnamefont {{Ren}}}, \bibinfo {author}
  {\bibfnamefont {C.}~\bibnamefont {{Wang}}}, \bibinfo {author} {\bibfnamefont
  {Y.}~\bibnamefont {{Zhang}}}, \bibinfo {author} {\bibfnamefont {Y.-H.}\
  \bibnamefont {{Hou}}}, \bibinfo {author} {\bibfnamefont {Y.-F.}\ \bibnamefont
  {{Wang}}},\ and\ \bibinfo {author} {\bibfnamefont {Z.-H.}\ \bibnamefont
  {{Cao}}},\ }\href {https://doi.org/10.1093/mnras/stw2096} {\bibfield
  {journal} {\bibinfo  {journal} {Monthly Notices of the Royal Astronomical
  Society}\ }\textbf {\bibinfo {volume} {463}},\ \bibinfo {pages} {2623}
  (\bibinfo {year} {2016})},\ \Eprint {https://arxiv.org/abs/1604.01216}
  {arXiv:1604.01216 [astro-ph.GA]} \BibitemShut {NoStop}%
\end{thebibliography}%

\clearpage
\onecolumngrid

\setcounter{page}{1}
\renewcommand{\thepage}{S\arabic{page}}

\setcounter{section}{0}
\setcounter{subsection}{0}
\setcounter{subsubsection}{0}
\setcounter{figure}{0}
\setcounter{table}{0}
\setcounter{equation}{0}

\setcounter{secnumdepth}{3}

\renewcommand{\thesection}{S\arabic{section}}
\renewcommand{\thesubsection}{\thesection.\arabic{subsection}}
\renewcommand{\thesubsubsection}{\thesubsection.\arabic{subsubsection}}

\renewcommand{\theequation}{S\arabic{equation}}
\renewcommand{\thefigure}{S\arabic{figure}}
\renewcommand{\thetable}{S\arabic{table}}

\begin{center}
{\large \bf Supplemental Material for}\\[0.5em]
{\large \bf ``Milky Way Dynamics Favor Dark Matter over Modified Gravity Models''}
\end{center}

\vspace{1em}

\section{Materials and Methods} 
\subsection{Construction of the rotation curve} 
\subsubsection{Data selection.} We utilize the high-quality Red Giant Branch (\textsc{RGB}) star sample from~\cite{Ou_2024}, 
which integrates \textsc{APOGEE DR17} spectroscopy~\cite{Majewski_2017} with multi-band photometry from Gaia DR3~\cite{Gaia_2023}, \textsc{2MASS}~\cite{2MASS}, and \textsc{WISE}~\cite{Wright_2010}. 
The final catalog consists of 120,309 \textsc{RGB} stars with complete 6D phase space information, 
and their spectrophotometric parallaxes offer a $\sim 40\%$ precision improvement over the original Gaia parallaxes. 
The chemical abundances are derived from \textsc{APOGEE DR17}, while the astrometric parameters, including Right Ascension, Declination, and proper motions, are sourced from Gaia DR3. We adopt the 6D cylindrical coordinate framework provided by~\cite{Ou_2024}.

\begin{figure}[!htbp]
    \centering
    \includegraphics[width=1\textwidth]{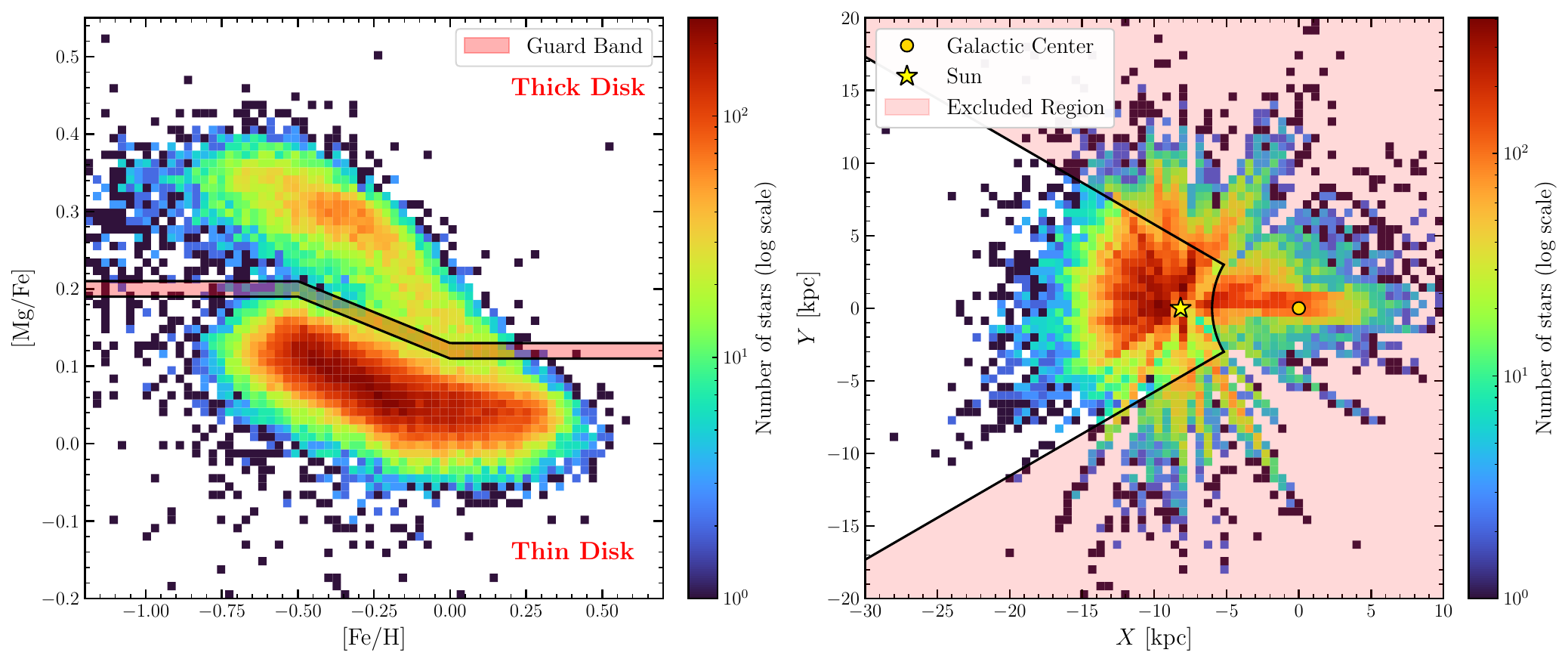}
    \caption{\textbf{Sample selection and spatial distribution of Red Giant Branch (RGB) stars.} \textbf{Left:} Distribution of the selected RGB sample in the chemical abundance plane of $[\mathrm{Mg/Fe}]$ versus $[\mathrm{Fe/H}]$. The color scale represents the stellar number density on a logarithmic scale. The red shaded region indicates the ``guard band'' used to separate the high-$\alpha$ thick disk and the low-$\alpha$ thin disk. The central demarcation line is similar to Adibekyan et al(2012)\cite{Adibekyan_2012}, defined as $y=0.20$ for $x<-0.5$, $y=-0.16x+0.12$ for $-0.5 \le x < 0$, and $y=0.12$ for $x \ge 0$ (where $x=[\mathrm{Fe/H}]$ and $y=[\mathrm{Mg/Fe}]$). A guard band with a width of $\pm 0.01$~dex is applied around this central line to minimize contamination. \textbf{Right:} Spatial distribution of the thin-disk sample in the Galactocentric $X-Y$ plane. The yellow star marks the Sun at $(X, Y) = (-8.178, 0)$~kpc, and the yellow circle indicates the Galactic Center. Solid black lines delineate the anti-center sector ($150^{\circ} \le \phi \le 210^{\circ}$) selected for analysis to mitigate the influence of non-axisymmetric structures and kinematic substructures. The red shaded area represents the excluded region, specifically requiring $R > 6$~kpc to mitigate the influence of the Galactic bar.
    \label{fig:selection}}
\end{figure}

We isolate a thin disk sample by a chemical cut in the $[\mathrm{Mg}/\mathrm{Fe}]-[\mathrm{Fe}/\mathrm{H}]$ plane~\cite{Adibekyan_2012}, incorporating a buffer zone between the thin and thick disk sequences to minimize contamination (see Fig.~\ref{fig:selection}). 
All kinematic calculations are performed in Galactocentric cylindrical coordinates $(R, \phi, z)$, adopting solar parameters consistent with Ou et al. (2024)~\cite{Ou_2024}, $R_\odot = 8.178$~kpc, $Z_\odot = 0.0208$~kpc, and a solar velocity vector $(v_x, v_y, v_z) = (5.1, 247.3, 7.8)~\mathrm{km~s^{-1}}$.

We limit the sample to the anti-center sector ($|\phi - 180^\circ| \le 30^\circ$) and select the near-plane region ($|z| < 1.0$~kpc) (see also Fig.~\ref{fig:selection}). 
We further apply a kinematic cut of $|v_z| < 100~\mathrm{km~s^{-1}}$ to exclude outliers and retain only those with velocity uncertainties below $10~\mathrm{km~s^{-1}}$. 
This selection yields a final sample of 40,101 stars for the rotation curve construction. 
Our final sample is then partitioned into radial bins spanning $R \in [6, 27.5]$~kpc. 
For each bin, the representative radius $R$ is defined as the average of the Galactocentric distances of the constituent stars, weighted by their uncertainties.

\subsubsection{Axisymmetric Jeans equation.} 
We derive the MW circular velocity $v_c(R)$ from the axisymmetric Jeans equation under the steady-state assumption~\cite{Binney_2008}. 
In cylindrical coordinates, the radial Jeans equation is
\begin{equation}
\frac{\partial(n_\star \langle v_R^2 \rangle)}{\partial R} + \frac{\partial(n_\star \langle v_R v_z \rangle)}{\partial z} + n_\star \left( \frac{\langle v_R^2 \rangle - \langle v_\phi^2 \rangle}{R} + \frac{\partial \Phi}{\partial R} \right) = 0,
\end{equation}
where $n_\star(R, z)$ is number density of the tracer stars. 
Since our stellar sample is strongly confined to the MW mid-plane ($|z| \approx 0$), we neglect the second term, the \textit{tilt term} that describes the misalignment of the velocity ellipsoid. 
The systematic bias introduced by this simplification, as well as deviations from the axisymmetric Jeans equation assumptions---such as non-axisymmetric effects and non-equilibrium dynamics---are accounted for during the fitting process by incorporating systematic errors following \cite{Ou_2025}.

\begin{figure}[p]
    \centering
    \includegraphics[width=1\textwidth]{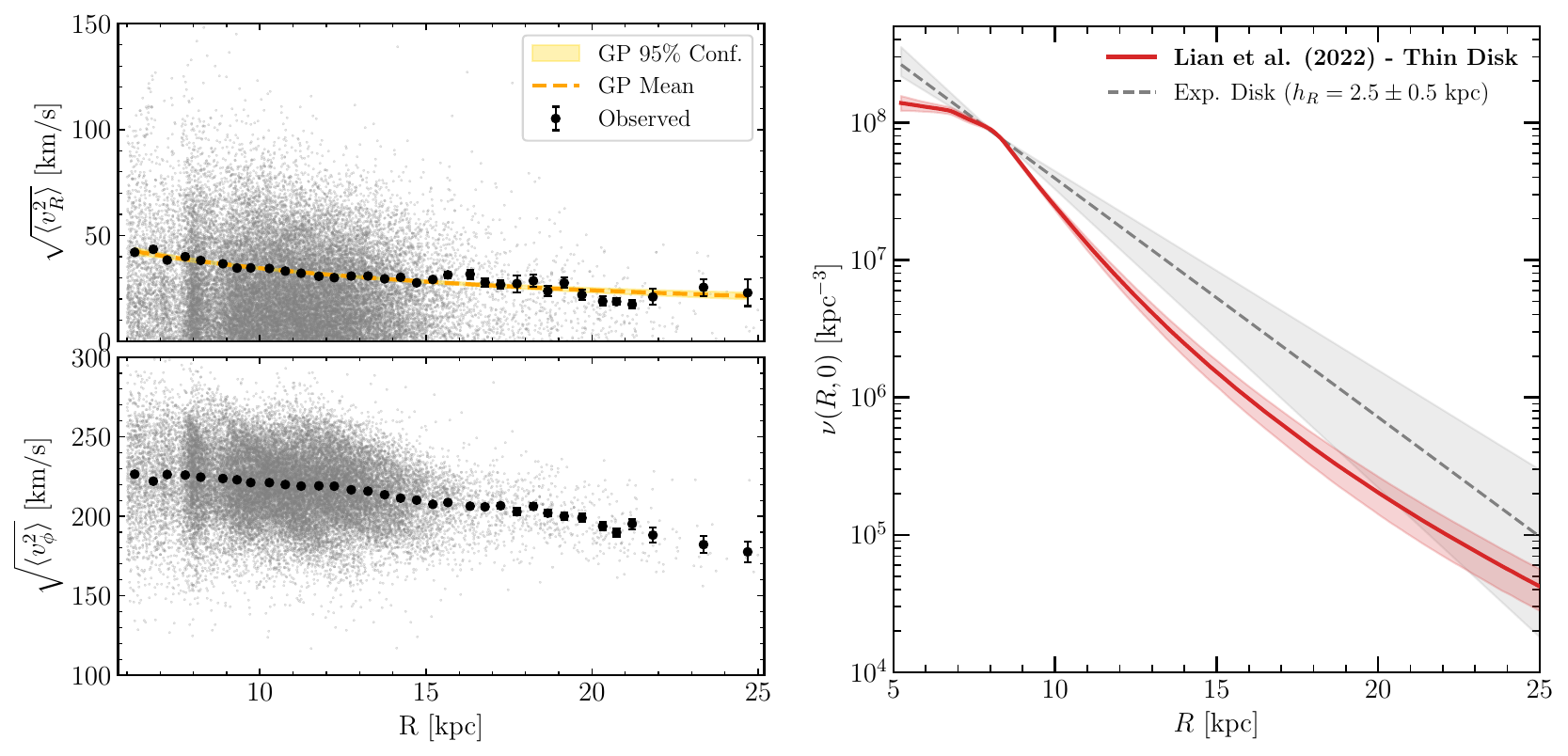}
    \caption{\textbf{Radial profiles of kinematic and density properties for the thin-disk sample.}\textbf{Left:} Radial distribution of the root-mean-square (RMS) velocities. The top and bottom panels show the radial RMS velocity $\sqrt{\langle v_R^2 \rangle}$ and the azimuthal RMS velocity $\sqrt{\langle v_\phi^2 \rangle}$, respectively. Black points with error bars represent the observational data binned in radius. The dashed orange line indicates the mean trend inferred from Gaussian Process (GP) regression, with the yellow shaded region denoting the 95\% confidence interval. \textbf{Right:} Mid-plane stellar number density profile $\nu(R, 0)$ of the thin disk. The red solid line represents the intrinsic number density distribution derived from the MAPs model of Lian et al. (2022)~\cite{Lian_2022}, which follows a broken-exponential profile. The red shaded area indicates the associated uncertainty. For comparison, the grey dashed line shows a single-exponential disk profile with a scale length of $h_R = 2.5 \pm 0.5$~kpc, illustrating the deviation of the traditional model from the actual complex structure of the MW disk.~\label{fig:ED2}}
\end{figure}

Substituting $v_c^2 = R\,(\partial \Phi/\partial R)$, the expression of the rotation curve becomes
\begin{equation}
v_c^2(R) = \langle v_\phi^2 \rangle - \langle v_R^2 \rangle \left(1 + \frac{\partial \ln n_\star}{\partial \ln R} + \frac{\partial \ln \langle v_R^2 \rangle}{\partial \ln R} \right).
\end{equation}
We utilize Gaussian Process (GP) regression~\cite{Aigrain_2023} to extract the velocity-dispersion gradient $\partial \ln \langle v_R^2 \rangle / \partial \ln R$ by modeling $\ln \langle v_R^2 \rangle$ as a function of $\ln R$. 
Our GP kernel combines a radial basis function with a white-noise term to filter local scatter and suppress fluctuations (see Fig.~\ref{fig:ED2}). 
The resulting gradient and its uncertainty are then derived via Monte Carlo sampling of the GP posterior.

The density gradient $\partial \ln n_\star / \partial \ln R$ for thin-disk RGB stars is taken from~\cite{Lian_2022}, 
which incorporates the \textsc{APOGEE} selection function. 
The model adopts a broken-exponential profile that is flat within the transition radius $R_b$ and exponential for $R > R_b$. 
We obtain the integrated RGB number density profile for the thin disk by summing all MAPs belonging to the thin disk sequence (see Fig.~\ref{fig:ED2}). 
This approach reduces systematic uncertainties present in earlier analyses that relied on single-exponential disk number density models, 
which poorly described the tracer population~\cite{Eilers_2019,Ou_2024,Zhou_2023}. 
The total rotation-curve uncertainty includes observational errors, Poisson noise, GP variance, and uncertainties in baryonic model parameters. as detailed in Table \ref{tab:rotation_data}.

\subsection{Reconstruction of the vertical potential from the phase‑space snail}

We reconstruct the MW vertical potential from the phase-space snail using the model-independent dynamical method developed by Guo et al.~\cite{Guo_2024} (hereafter G24). 
This approach avoids the equilibrium assumption required by traditional Jeans-equation analyses~\cite{Buch_2019,Guo_2020,Cheng_2024,Soding_2025}. 
The snail itself is a direct signature of non-equilibrium in the MW disk, and its structure encodes the non-equilibrium dynamical information needed to recover the vertical potential without steady-state assumptions.

Our implementation closely follows the methodology of G24, with the primary refinement being the use of our independently constructed MW rotation curve to map angular momentum $L_z$ to the guiding-center radius $R_g$. 
We utilize the Gaia DR3 catalog with radial velocities and Bayesian distances~\cite{Bailer_2021}, adopting sample selection criteria consistent with G24.

Assuming the decoupling of vertical and in-plane motions, the phase-space ``snail'' structure traces isolines of vertical energy $E_z$. 
The intersections of these structures with the $z$- and $V_z$-axes correspond to the turning points $Z_{\mathrm{max}}$ and mid-plane crossings $V_{z,\mathrm{max}}$, respectively, satisfying $E_z = \Phi_z(Z_{\mathrm{max}}) = V_{z,\mathrm{max}}^2 / 2$. 
Following G24, we linearly interpolate between adjacent intersections to recover the full potential-energy pairs $(Z_{\mathrm{max}}, \Phi_z)$ without relying on specific density models.

To mitigate radial mixing, stars are binned by $R_g$. 
We apply an empirical correction for asymmetric drift to map the observed $R_g$ to an effective potential radius $R_{\mathrm{pot}}$. 
Phase-space intersections are identified from density-contrast maps via Gaussian fitting. 
This procedure allows us to reconstruct the vertical potential across the range $7.8~{\rm kpc} < R_{\mathrm{pot}} < 11.0~{\rm kpc}$ (see Table \ref{tab:vertical_potential_data}), with 
a relative accuracy better than $5\%$ \cite{Guo_2024}.
Because this method essentially treats the perturbed stars as massless test particles, the current theoretical framework does not capture the complex dynamical response arising from the self-gravity of the spiral structure itself. Although including self-gravity in a non-linear theory would alter the detailed density contrasts within the snails, the overall angles and periods of these structures are dictated by the background potential to zeroth order. Consequently, the Galactic vertical force inferred from our test-particle approximation remains robust, particularly after incorporating conservative errors. To further complete the theoretical picture, the self-gravity of the phase-space spiral remains an important avenue for future detailed modeling.

\subsection{Theoretical Frameworks: Dark Matter Halos, MOND, and STVG}

In the axisymmetric cylindrical coordinate system $(R, z)$ adopted in this study, 
the total gravitational potential $\Phi(R, z)$ is determined by the chosen gravitational framework. 
The circular velocity $v_c(R)$ and the vertical potential $\Phi_z(R, z)$ are derived from the radial and vertical gradients of the total potential,
\begin{equation}
v_c^2(R) = R \left. \frac{\partial \Phi}{\partial R} \right|_{z=0}, \quad {\rm where}~~\Phi_{z}(R,z) = \Phi(R, z) - \Phi(R, 0).
\end{equation}

\subsubsection{Dark Matter models.} The total gravitational potential is the sum of the contributions from baryonic matter and a nearly spherical dark matter halo,
\begin{equation}
\Phi(R, z) = \Phi_{\text{baryon}}(R, z) + \Phi_{\text{halo}}(R, z) .
\end{equation}
The corresponding Poisson equation is
\begin{equation}
\nabla^2 \Phi_{\text{tot}} = 4\pi G (\rho_{\text{baryon}} + \rho_{\text{halo}}).
\end{equation}
We model the dark matter halo with either the NFW profile~\cite{Navarro_1997} or the Einasto profile~\cite{Einasto_1965}:
\begin{equation}
\rho_{\text{NFW}}(r) = \frac{\rho_s}{(r/r_s)(1 + r/r_s)^2}, \qquad 
\rho_{\text{Ein}}(r) = \rho_s \exp \left\{ -\frac{2}{\alpha_{\text{Ein}}} \left[ \left(\frac{r}{r_s}\right)^{\alpha_{\text{Ein}}} - 1 \right] \right\}.
\end{equation}
We parameterize the profiles using the local dark matter density at the solar position $\rho_{\odot,\text{DM}} \equiv \rho(R_{\odot})$ rather than the scale density $\rho_s$, where $R_{\odot} = 8.178$~kpc.

With $\rho_s$ fixed by $\rho_{\odot,\text{DM}}$, the halo radius $R_{200}$ and mass $M_{200}$ are obtained by solving:
\begin{equation}
M_{200} = \int_0^{R_{200}} 4\pi r^2 \rho(r) \, \mathrm{d}r = \frac{4}{3}\pi R_{200}^3 (200 \rho_{\text{crit}}),
\end{equation}
where $\rho_{\text{crit}}$ is the critical density of the universe. The concentration parameter is then simply defined as~\cite{Dutton_2014}:
\begin{equation}
c_{200} = \frac{R_{200}}{r_s}.
\label{c}
\end{equation}
Since $r_s$ corresponds to the radius where the logarithmic density slope is $-2$ in both models, this definition provides a consistent comparison.

We adopt flat priors on the halo parameters: $\log_{10}(\rho_{\odot,\text{DM}}/\text{GeV\,cm}^{-3}) \in [-3.0, 1.0]$ and $\log_{10}(r_s/\text{kpc}) \in [0.0, 2.3]$. 
For the Einasto profile the shape parameter $\alpha_{\text{Ein}}$ is also given by a flat prior in $[0,1]$. 

\subsubsection{Modified Gravity Theories.} In modified gravity frameworks such as MOND and STVG, 
the gravitational field exhibits a non-linear response to the baryonic matter distribution. 

For QUMOND~\cite{Milgrom_1983, Milgrom_2010}, we employ the phantom dark matter approach~\cite{Famaey_2012}. 
In this framework, the excess gravity is mathematically attributed to a fictitious ``phantom'' mass density that is completely determined by the baryonic distribution.
We first derive the Newtonian potential $\Phi_N$ via the Poisson equation $\nabla^2 \Phi_N = 4\pi G \rho_{\text{baryon}}$ and subsequently compute the phantom dark matter density $\rho_p$ as
\begin{equation}
\rho_p = \frac{1}{4\pi G} \nabla \cdot \left[ \left( \nu\left(\frac{|\nabla \Phi_N|}{a_0}\right) - 1 \right) \nabla \Phi_N \right] .
\end{equation}
The total gravitational potential is then determined by $\nabla^2 \Phi = 4\pi G (\rho_{\text{baryon}} + \rho_p)$. 
In our analysis, the acceleration scale is fixed at $a_0 = 1.2 \times 10^{-10}~\mathrm{m\,s^{-2}}$. 
However, for the robustness tests, we relax it and adopt a flat prior for $a_0$ within the range of $[0.3, 3.0] \times 10^{-10}~\mathrm{m\,s^{-2}}$.
This broad interval is designed to extensively cover the parameter space explored in previous studies, ensuring that our analysis is not restricted to the standard value.
We evaluate three standard interpolation functions $\nu(y)$ \cite{Famaey_2012, McGaugh_2016}, including
\begin{equation}
\nu_{\textbf{Simple}}(y) = \frac{1}{2} + \sqrt{\frac{1}{4} + \frac{1}{y}},
\nu_{\textbf{Standard}}(y) = \sqrt{\frac{1}{2} \left( 1 + \sqrt{1 + \frac{4}{y^2}} \right)} ,
\nu_{\textbf{RAR}}(y) = \frac{1}{1 - e^{-\sqrt{y}}}.
\end{equation}

To mitigate the significant baryon shortfall in the inner MW, we enhance the phantom matter distribution in the transition regime ($g_{\text{obs}} \approx a_0$) by generalizing the classic exponential $\mu$-interpolation function (e.g., \cite{Zhao_2007}) into the universal $E$-$\alpha$ family:
\begin{equation}
    \mu_\alpha(x) = \left[ 1 - \exp\left( -x^\alpha \right) \right]^{1/\alpha}
    \label{eq:mu_alpha}
\end{equation}
where $x = g_{\text{obs}}/a_0$. Within the QUMOND framework, the corresponding $\nu$-function is defined by $\nu_\alpha(y) = 1/\mu_\alpha(x)$ via the  algebraic mapping $x = y\nu_\alpha(y)$, with $y = g_{\mathrm{bar}}/a_0$. 
A salient advantage of this exponential generalization is its rapid convergence to the Newtonian limit ($\nu_\alpha \to 1$) in the strong-field regime. 
However, as demonstrated in Fig.~3, even with such high flexibility in the interpolation function, MOND still exhibits inherent limitations in reconciling the multi-component kinematic data of the MW.

In STVG~\cite{Moffat_2006}, 
the effective potential is composed of an enhanced Newtonian component and a repulsive Yukawa term
\begin{equation}
\Phi_{\text{STVG}}(R, z) = (1+\alpha_{\text{STVG}}) \Phi_N(R, z) + \Phi_Y(R, z) .
\end{equation}
The Yukawa potential $\Phi_Y$ is governed by 
$(\nabla^2 - \mu_{\text{STVG}}^2) \Phi_Y = -4\pi G \alpha_{\text{STVG}} \rho_{\text{baryon}}$. 
Following~\cite{Davari_2020}, we adopted flat priors of $\alpha_{\text{STVG}} \in [7.0, 16.0]$ and $\mu_{\text{STVG}} \in [0.03, 0.07]~\mathrm{kpc}^{-1}$. This choice allows the model maximum flexibility to specifically fit the MW dynamics.

Numerical potentials are computed by solving the Poisson equation on a $40 \times 40$ kpc cylindrical grid. While the radial grid is uniform, the vertical grid spacing increases geometrically with $|z|$ to resolve steep density gradients near the mid-plane. We employ a standard five-point finite-difference scheme solved via direct matrix decomposition, with boundary conditions determined by multipole expansion.
Within our primary region of interest, numerical errors are maintained below $1\%$.

\subsection{Baryonic Mass Distribution}

We follow the conventional decomposition of the MW baryonic mass into three components: 
a stellar disk, a gas disk, and a bulge. 
Instead of a simple exponential, we adopt a broken-exponential radial profile for the stellar disk, 
as proposed by \cite{Lian_2025}, which better matches the observed surface density in the inner disk.

For the total baryonic mass model we adopt the comprehensive stellar distribution~\cite{Lian_2022}.
The stellar component is represented as a thin‑ plus a thick‑disk, each following a broken‑exponential radial profile. 
The two‑dimensional density of a disk layer is
\begin{equation}
\rho_{\text{disk}}(R, z) = \frac{\Sigma(R)}{2 h_z(R)} \exp\left[-\frac{|z|}{h_z(R)}\right],
\end{equation}
with a surface density that breaks at radius $R_b$
\begin{equation}
\Sigma(R) = 
\begin{cases} 
\Sigma_{b} \exp\left[ -(R - R_b)/h_{\text{inner}} \right], & R \le R_b \\ 
\Sigma_{b} \exp\left[ -(R - R_b)/h_{\text{outer}} \right], & R > R_b.
\label{eq:sigma_b}
\end{cases}
\end{equation}
The vertical scale height flares outward as $h_z(R) = h_{z,\odot} \exp\bigl[ A_{\text{flare}} (R - R_\odot) \bigr]$. 
We fix the geometric parameters to the values given by \cite{Lian_2022}. 
Specifically, we adopt for the thin disk $R_b = 7.46$~kpc, $h_{\text{outer}} = 2.08$~kpc, $h_{z,\odot} = 0.39$~kpc, and $A_{\text{flare}} = 0.027$; and for the thick disk $R_b = 7.31$~kpc, $h_{\text{outer}} = 1.47$~kpc, $h_{z,\odot} = 0.85$~kpc, and $A_{\text{flare}} = 0.057$. 
The inner profiles ($R \le R_b$) for both disks are taken to be flat ($h_{\text{inner}} \to \infty$).

In our dynamical model, the local normalized density coefficients $\rho_{\odot,\text{thin}}$ and $\rho_{\odot,\text{thick}}$ serve as the free parameters. 
Their fiducial values are based on the local stellar density derived from \textit{Gaia} DR3 by \cite{Lian_2025}. 
Following \cite{McKee_2015}, we apply a conversion factor of 1.25 to include stellar remnants in the total stellar density. 
To account for uncertainty in this factor, we conservatively increase the original uncertainties by 50\%. 
Finally, we assign Gaussian priors to both parameters, i.e.,  $\rho_{\odot,\text{thin}} \sim \mathcal{N}(0.045, 0.003)$ and $\rho_{\odot,\text{thick}} \sim \mathcal{N}(0.00525, 0.0006)$, in units of $M_\odot\,\mathrm{pc}^{-3}$.

The gas disk and bulge are included as fixed components. 
The gas disk follows the density distribution~\cite{McMillan_2016}
\begin{equation}
\rho_{\mathrm{gas}}(R, z) = \frac{\Sigma_0}{4 z_d} \exp \left( -\frac{R_m}{R} - \frac{R}{R_d} \right) \mathrm{sech}^2 \left( \frac{z}{2 z_d} \right),
\label{eq:gas_den}
\end{equation}
which includes both atomic (HI) and molecular (H$_2$) hydrogen. The adopted parameters are: for HI, $\Sigma_0 = 53.1\,M_\odot\,\mathrm{pc}^{-2}$, $R_d = 7.0$~kpc, $R_m = 4.0$~kpc, and $z_d = 0.085$~kpc; 
for H$_2$, $\Sigma_0 = 2180\,M_\odot\,\mathrm{pc}^{-2}$, $R_d = 1.5$~kpc, $R_m = 12.0$~kpc, and $z_d = 0.045$~kpc. At the solar radius, this model yields a surface density of $\Sigma_{\mathrm{HI},\odot} \approx 11.0 \, M_\odot \, \mathrm{pc}^{-2}$ and $\Sigma_{\mathrm{H}_2,\odot} \approx 2.0 \, M_\odot \, \mathrm{pc}^{-2}$.

For the bulge we hire a spherically symmetric two-component Multi-Gaussian Expansion model
\begin{equation}
\rho_{\mathrm{bulge}}(r) = \sum_{i=1}^{2} \frac{M_i}{(2\pi)^{3/2} \sigma_i^3} \exp \left( -\frac{r^2}{2\sigma_i^2} \right),
\end{equation}
with $M_1 = 6.5 \times 10^9\,M_\odot$, $\sigma_1 = 0.5$~kpc, $M_2 = 1.48 \times 10^{10}\,M_\odot$, and $\sigma_2 = 1.4$~kpc. 
This spherical approximation reproduces the root‑mean‑square rotation curve of the non‑axisymmetric bar model~\cite{Sormani_2022} within $\pm 30^\circ$ of the Sun–Galactic centre line. 
Over the radial range of interest, the systematic error introduced by this approximation remains below $5\,\mathrm{km\,s^{-1}}$.

\begin{figure}[p]
    \centering
    \includegraphics[width=1\textwidth]{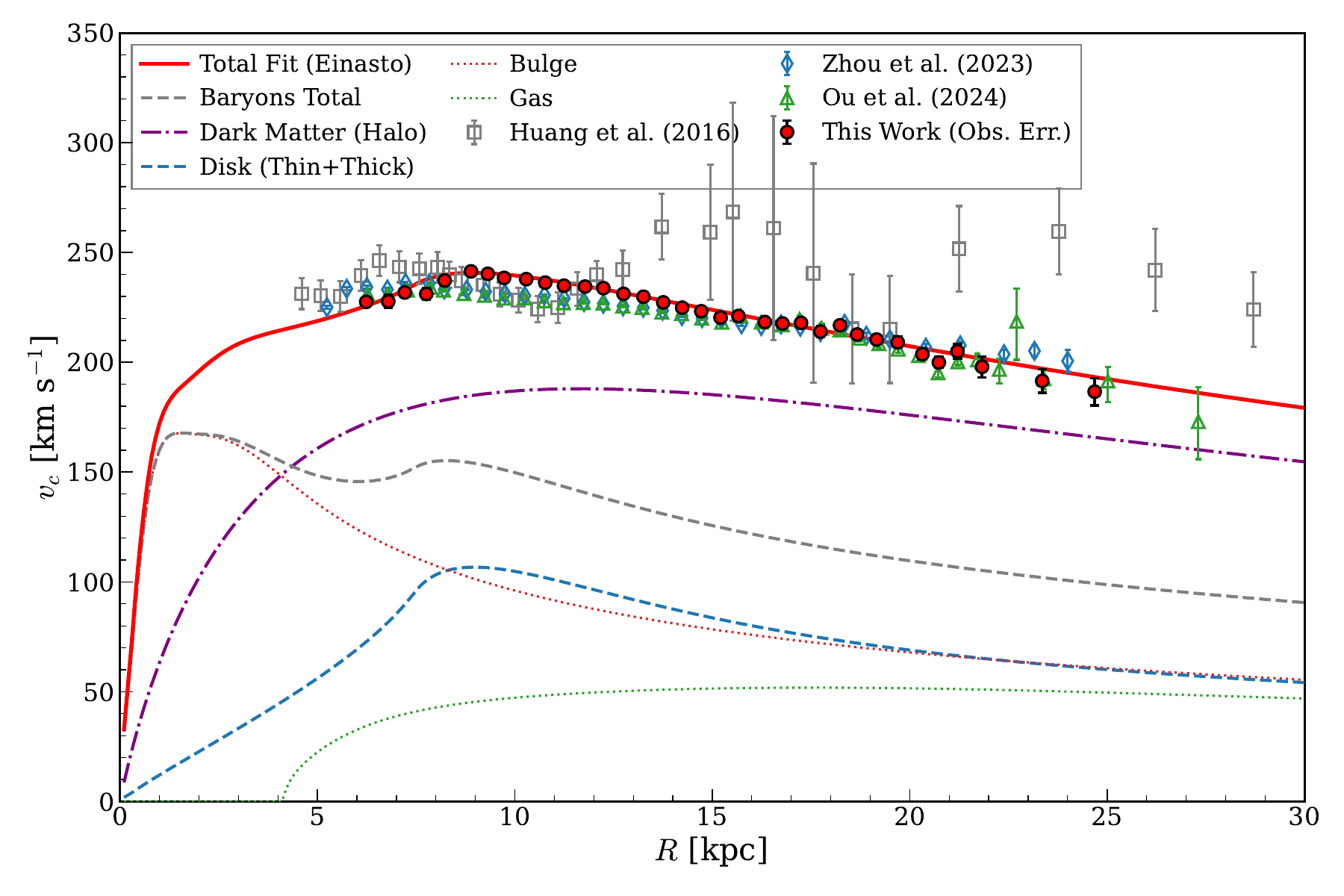}
    \caption{
    \textbf{Decomposition of the MW rotation curve for the best-fit Einasto dark matter halo model.} The red circles with error bars represent the high-precision rotation curve derived in this work with observational error. Compared to other rotation curves, our results exhibit a distinct ``bump'' feature around $R \approx 7-12$~kpc. This feature is precisely explained by the contribution from the broken-exponential (truncated) stellar disk (blue dashed line). For comparison, open symbols denote measurements from other studies: grey squares from Huang et al.~(2016)~\cite{Huang_2016}, blue diamonds from Zhou et al.~(2023)~\cite{Zhou_2023}, and green triangles from Ou et al.~(2024)~\cite{Ou_2024}.
    \label{fig:ED4}}
\end{figure}

The resulting mass decomposition is presented in Fig.~\ref{fig:ED4}, which also compares the high-precision rotation curve derived in this work with other measurements. By employing the broken-exponential profile to describe the tracer density, our constructed rotation curve reveals a distinct ``bump'' feature. Remarkably, the gravitational contribution of the adopted broken-exponential stellar disk generates a corresponding rise, thereby self-consistently explaining the observed ``bump'' structure.

Finally, the governing equations and prior ranges for all models are summarized in Table~\ref{tab:models_priors}.

\subsection{Statistical Inference}

We use a Bayesian framework to obtain the goodness of models and perform the model comparision. 
The posterior probability distribution $P(\boldsymbol{\Theta} \mid \mathcal{D})$ of the parameter set $\boldsymbol{\Theta}$ follows Bayes' theorem
\begin{equation}
P(\boldsymbol{\Theta} \mid \mathcal{D}) = \frac{\mathcal{L}(\mathcal{D} \mid \boldsymbol{\Theta})\, \pi(\boldsymbol{\Theta})}{\mathcal{Z}},
\end{equation}
where the prior probability distributions of prior and likelihood are $\pi(\boldsymbol{\Theta})$ and $\mathcal{L}(\mathcal{D} \mid \boldsymbol{\Theta})$, respectively.  
The information for model comparison is encoded in the Bayesian evidence $\mathcal{Z}$ (see, e.g., \cite{Trotta_2008,Burnham_2002,Gelman_2013}).

The total log‑likelihood is the sum of contributions from the rotation curve $v_c$ and the vertical gravitational potential $\Phi_z$
\begin{equation}
\ln \mathcal{L}_{\mathrm{total}}(\boldsymbol{\Theta}) = \ln \mathcal{L}_{\mathrm{rot}} + \ln \mathcal{L}_{\mathrm{vert}}.
\end{equation}

For the rotation curve with $N_{\mathrm{rot}}$ data points, the log-likelihood function is given by
\begin{equation}
\ln \mathcal{L}_{\mathrm{rot}} = -\frac{1}{2} \sum_{i=1}^{N_{\mathrm{rot}}} \left[ \frac{\bigl(v_{c,\mathrm{obs}}(R_i) - v_{c,\mathrm{mod}}(R_i, \boldsymbol{\Theta})\bigr)^2}{\sigma_{v,i}^2} + \ln\bigl(2\pi \sigma_{v,i}^2\bigr) \right], 
\label{eq:Lrot}
\end{equation}
where $N_{\mathrm{rot}}$ is the total number of measurements. The quantities $v_{c,\mathrm{obs}}(R_i)$ and $\sigma_{v,i}$ represent the observed circular velocity and its corresponding uncertainty at the radial distance $R_i$, respectively. The term $v_{c,\mathrm{mod}}(R_i, \boldsymbol{\Theta})$ denotes the model-predicted velocity given the set of free parameters $\boldsymbol{\Theta}$.

For the vertical potential, we account for the uncertainty in both the potential and the vertical coordinate $z$. The total error is
\begin{equation}
\sigma_{\mathrm{eff},j,k}^2 = \sigma_{\Phi,j,k}^2 + \left( \left. \frac{\partial \Phi_{z,\mathrm{mod}}}{\partial z} \right|_{z_j,R_k} \cdot \sigma_{z,j,k} \right)^2 = \sigma_{\Phi,j,k}^2 + \bigl[ K_z(z_j,R_k) \cdot \sigma_{z,j,k} \bigr]^2,
\end{equation}
with \(K_z = -\partial \Phi_z / \partial z\) the vertical restoring force. The corresponding log‑likelihood is
\begin{equation}
\ln \mathcal{L}_{\mathrm{vert}} = -\frac{1}{2} \sum_{k} \sum_{j=1}^{N_{z,k}} \left[ \frac{\bigl(\Phi_{z,\mathrm{obs}}(z_j,R_k) - \Phi_{z,\mathrm{mod}}(z_j,R_k ,\boldsymbol{\Theta})\bigr)^2}{\sigma_{\mathrm{eff},j,k}^2} + \ln\bigl(2\pi \sigma_{\mathrm{eff},j,k}^2\bigr) \right].\label{eq:Lvert}
\end{equation}

We scan the parameter space with the MCMC algorithm implemented in the \texttt{emcee} package~\cite{Foreman_2013}. 
With checked convergence, each model is run with 32 walkers for 20,000 steps, discarding the first 6,000 as burn‑in. 

For model comparison, we compute the Bayesian evidence $\mathcal{Z}$ using a grid integration method. 
The evidence is defined as the marginal likelihood over the entire prior space:
\begin{equation}
\mathcal{Z} = \int \mathcal{L}(\mathcal{D} \mid \boldsymbol{\Theta}) \, \pi(\boldsymbol{\Theta}) \, d\boldsymbol{\Theta}.
\end{equation}
Since the posterior distributions are well confined, integrating over the high‑likelihood region with a regular grid yields an accurate estimate of $\ln\mathcal{Z}$.

Model selection is primarily based on the Bayesian Information Criterion (BIC)~\cite{Schwarz_1978}:
\begin{equation}
\mathrm{BIC} = -2 \ln\hat{\mathcal{L}} + k \ln N,
\end{equation}
where $\hat{\mathcal{L}}$ is the maximum likelihood, $k$ the number of free parameters, and $N$ the total number of data points. 
The BIC penalises model complexity and provides a convenient approximation to the fully Bayesian evidence. 
We also report the reduced chi‑square,
\begin{equation}
\chi^2_{\mathrm{red}} = \frac{1}{N-k} \sum_{i=1}^{N} \frac{(O_i - E_i)^2}{\sigma_i^2},
\end{equation}
as a measure of goodness‑of‑fit. 

Following Jeffreys’ scale, a difference $\Delta\mathrm{BIC}>10$ between two models indicates \textit{strong evidence against the model with the higher BIC}~\cite{Burnham_2002}. 
To quantify the failure of each alternative model relative to the best‑performing one, 
we compute the probability $P(M_{\mathrm{model}} \mid D)$ and convert it to an equivalent significance in unit of Gaussian half width $\sigma$ as 
\begin{equation}
P(M_{\mathrm{model}} \mid D) = \frac{1}{1 + e^{\Delta \ln\mathcal{Z}}}, \qquad {\rm and}~~ 
{\rm significance} = \sqrt{2} \, \mathrm{erf}^{-1}\!\bigl[1 - P(M_{\mathrm{model}} \mid D)\bigr],
\end{equation}
with $\Delta \ln\mathcal{Z} = \ln\mathcal{Z}_{\mathrm{best}} - \ln\mathcal{Z}_{\mathrm{model}}$. 

All fitting statistics and the fitted values of the parameters are presented in Table~\ref{tab:comprehensive_fits_final}.

\section{Supplementary Text}

\textbf{Full Acknowledgments for Survey Data:}
Funding for the Sloan Digital Sky Survey IV has been provided by the Alfred P. Sloan Foundation, the U.S. Department of Energy Office of Science, and the Participating Institutions. SDSS-IV acknowledges support and resources from the Center for High-Performance Computing at the University of Utah. The SDSS web site is www.sdss.org. SDSS-IV is managed by the Astrophysical Research Consortium for the Participating Institutions of the SDSS Collaboration including the Brazilian Participation Group, the Carnegie Institution for Science, Carnegie Mellon University, the Chilean Participation Group, the French Participation Group, Harvard-Smithsonian Center for Astrophysics, Instituto de Astrofísica de Canarias, The Johns Hopkins University, Kavli Institute for the Physics and Mathematics of the Universe (IPMU) / University of Tokyo, the Korean Participation Group, Lawrence Berkeley National Laboratory, Leibniz Institut für Astrophysik Potsdam (AIP), Max-Planck-Institut für Astronomie (MPIA Heidelberg), Max-Planck-Institut für Astrophysik (MPA Garching), Max-Planck-Institut für Extraterrestrische Physik (MPE), National Astronomical Observatories of China, New Mexico State University, New York University, University of Notre Dame, Observatário Nacional / MCTI, The Ohio State University, Pennsylvania State University, Shanghai Astronomical Observatory, United Kingdom Participation Group, Universidad Nacional Autónoma de México, University of Arizona, University of Colorado Boulder, University of Oxford, University of Portsmouth, University of Utah, University of Virginia, University of Washington, University of Wisconsin, Vanderbilt University, and Yale University.

This publication makes use of data products from the Two Micron All Sky Survey, which is a joint project of the University of Massachusetts and the Infrared Processing and Analysis Center/California Institute of Technology, funded by the National Aeronautics and Space Administration and the National Science Foundation.

This publication also makes use of data products from the Wide-field Infrared Survey Explorer, which is a joint project of the University of California, Los Angeles, and the Jet Propulsion Laboratory/California Institute of Technology, funded by the National Aeronautics and Space Administration.

\clearpage 

\clearpage

\clearpage
\clearpage

\begin{table}
\centering
\caption{\textbf{Numerical data of the MW rotation curve.} This table provides the binned Galactocentric radius ($R_{\text{center}}$), the derived circular velocity ($v_c$), the observational uncertainty ($\sigma_{\text{obs}}$), and the total uncertainty ($\sigma_{\text{total}}$, incorporating both observational and systematic errors) used in our dynamical analysis.}
\label{tab:rotation_data}
\footnotesize
\renewcommand{\arraystretch}{0.8}
\begin{tabular}{cccc}

\hline
$R_{\text{center}}$ (kpc) & $v_c$ (km s$^{-1}$) & $\sigma_{\text{obs}}$ (km s$^{-1}$) & $\sigma_{\text{total}}$ (km s$^{-1}$) \\
\hline
6.23  & 227.56 & 1.77 & 13.73 \\
6.79  & 227.74 & 3.05 & 14.25 \\
7.21  & 231.82 & 1.53 & 14.48 \\
7.75  & 231.14 & 2.68 & 14.91 \\
8.22  & 237.28 & 1.66 & 15.44 \\
8.89  & 241.44 & 1.06 & 16.11 \\
9.31  & 240.33 & 1.09 & 16.34 \\
9.73  & 238.46 & 1.22 & 16.53 \\
10.29 & 237.90 & 1.32 & 16.94 \\
10.77 & 236.31 & 1.44 & 17.24 \\
11.24 & 234.92 & 1.37 & 17.54 \\
11.78 & 234.51 & 1.36 & 18.00 \\
12.24 & 233.80 & 1.32 & 18.39 \\
12.75 & 231.21 & 1.24 & 18.69 \\
13.26 & 229.88 & 1.15 & 19.11 \\
13.76 & 227.32 & 1.16 & 19.45 \\
14.24 & 224.91 & 1.15 & 19.80 \\
14.72 & 223.29 & 1.19 & 20.24 \\
15.21 & 220.41 & 1.29 & 20.60 \\
15.66 & 221.13 & 1.33 & 21.27 \\
16.33 & 218.37 & 1.61 & 21.96 \\
16.78 & 217.71 & 1.59 & 22.55 \\
17.25 & 217.99 & 1.97 & 23.34 \\
17.74 & 214.01 & 2.49 & 23.75 \\
18.23 & 216.89 & 1.88 & 24.87 \\
18.67 & 212.70 & 1.99 & 25.17 \\
19.16 & 210.39 & 2.19 & 25.82 \\
19.70 & 209.15 & 2.54 & 26.74 \\
20.32 & 203.71 & 2.52 & 27.31 \\
20.74 & 199.96 & 2.54 & 27.69 \\
21.21 & 205.05 & 3.24 & 29.52 \\
21.83 & 197.98 & 4.79 & 30.16 \\
23.36 & 191.53 & 5.37 & 33.10 \\
24.68 & 186.64 & 6.29 & 36.22 \\
\hline
\end{tabular}
\end{table}

\begin{table}[htbp]
\caption{\textbf{Numerical data of the vertical gravitational potential.} This table lists the reconstructed vertical potential $\Phi_z$ and the corresponding vertical height $|z|$ at different effective potential radii $R_{\rm pot}$. The uncertainties $\sigma_z$ and $\sigma_{\Phi_z}$ are placed immediately following their respective variables.}
\label{tab:vertical_potential_data}
\footnotesize
\setlength{\tabcolsep}{3.5pt}
\renewcommand{\arraystretch}{1.1} 
\begin{tabular}{ccccc | ccccc}
\hline
$R_{\rm pot}$ & $|z|$ & $\sigma_z$ & $\Phi_z$ & $\sigma_{\Phi_z}$ & $R_{\rm pot}$ & $|z|$ & $\sigma_z$ & $\Phi_z$ & $\sigma_{\Phi_z}$ \\
(kpc) & (kpc) & (kpc) & (km$^2$ s$^{-2}$) & (km$^2$ s$^{-2}$) & (kpc) & (kpc) & (kpc) & (km$^2$ s$^{-2}$) & (km$^2$ s$^{-2}$) \\
\hline
7.79 & 0.300 & 0.019 & 180.5 & 28.5 &  9.14 & 0.387 & 0.018 & 249.9 & 33.5 \\
7.79 & 0.421 & 0.025 & 311.8 & 26.5 &  9.14 & 0.474 & 0.025 & 360.6 & 28.5 \\
7.79 & 0.516 & 0.018 & 478.8 & 46.4 &  9.14 & 0.583 & 0.018 & 491.6 & 47.0 \\
7.79 & 0.621 & 0.025 & 568.9 & 35.8 &  9.14 & 0.704 & 0.025 & 638.0 & 37.9 \\
7.79 & 0.726 & 0.018 & 666.7 & 54.8 &  9.14 & 0.791 & 0.018 & 803.4 & 60.1 \\
7.79 & 0.840 & 0.025 & 908.8 & 45.2 &  9.54 & 0.367 & 0.018 & 222.2 & 31.6 \\
7.79 & 0.944 & 0.018 & 1188.3 & 73.1 &  9.54 & 0.420 & 0.025 & 298.3 & 25.9 \\
7.79 & 1.054 & 0.025 & 1269.3 & 53.4 &  9.54 & 0.539 & 0.018 & 385.6 & 41.7 \\
8.17 & 0.292 & 0.019 & 160.3 & 26.9 &  9.54 & 0.673 & 0.025 & 540.3 & 34.9 \\
8.17 & 0.414 & 0.025 & 288.5 & 25.5 &  9.88 & 0.283 & 0.018 & 133.8 & 24.5 \\
8.17 & 0.524 & 0.018 & 454.2 & 45.2 &  9.88 & 0.350 & 0.025 & 186.5 & 20.5 \\
8.17 & 0.645 & 0.025 & 568.8 & 35.8 &  9.88 & 0.488 & 0.018 & 248.0 & 33.4 \\
8.17 & 0.718 & 0.018 & 696.2 & 56.0 &  9.88 & 0.649 & 0.025 & 450.7 & 31.8 \\
8.17 & 0.795 & 0.025 & 876.6 & 44.4 & 10.24 & 0.320 & 0.025 & 135.0 & 17.4 \\
8.17 & 0.921 & 0.018 & 1077.7 & 69.6 & 10.24 & 0.444 & 0.018 & 195.9 & 29.7 \\
8.17 & 1.055 & 0.025 & 1294.8 & 54.0 & 10.24 & 0.587 & 0.025 & 430.4 & 31.1 \\
8.66 & 0.289 & 0.025 & 152.9 & 18.5 & 10.64 & 0.304 & 0.025 & 118.9 & 16.4 \\
8.66 & 0.379 & 0.018 & 240.6 & 32.9 & 10.64 & 0.425 & 0.018 & 177.6 & 28.3 \\
8.66 & 0.481 & 0.025 & 365.2 & 28.7 & 10.64 & 0.566 & 0.025 & 385.6 & 29.5 \\
8.66 & 0.586 & 0.018 & 515.6 & 48.2 & 10.95 & 0.296 & 0.025 & 108.9 & 15.7 \\
8.66 & 0.702 & 0.025 & 653.6 & 38.3 & 10.95 & 0.415 & 0.018 & 156.8 & 26.6 \\
9.14 & 0.309 & 0.025 & 143.3 & 18.0 & 10.95 & 0.553 & 0.025 & 333.2 & 27.4 \\
\hline
\end{tabular}
\end{table}
\clearpage
\begin{table}
    \centering
    \footnotesize 
    \caption{\textbf{Summary of gravitational models, governing equations, free parameters, and prior distributions used in the Bayesian inference.} $\mathcal{N}(\mu, \sigma)$ denotes a Gaussian prior and $\mathcal{U}[a, b]$ denotes a flat prior.}
    \label{tab:models_priors}
    \renewcommand{\arraystretch}{1.5} 
    \begin{tabular}{lp{6cm}p{3.5cm}p{3cm}}
        \hline
        Model/Component & Governing Equation / Profile & Parameter [Unit] & Prior \\
        \hline
        \multicolumn{4}{c}{\textbf{Baryonic Components}} \\
        \hline
        Stellar Disks & 
        $\begin{aligned} 
            & \rho(R,z) = \frac{\Sigma(R)}{2h_z(R)} \exp\left(-\frac{|z|}{h_z(R)}\right) \\ 
            & h_z(R) = h_{z,\odot} \exp\left[A_{\text{flare}}(R-R_{\odot})\right]
        \end{aligned}$ 
        & $\rho_{\odot,\text{thin}}$\newline $\rho_{\odot,\text{thick}}$\newline $[M_{\odot}\text{pc}^{-3}]$ 
        & $\mathcal{N}(0.045, 0.003)$ \newline $\mathcal{N}(0.00525, 0.0006)$ \newline (Shape params fixed) \\ 
        
        Bulge & $\rho_{\text{b}}(r)=\sum_{i=1}^{2}\frac{M_{i}}{(2\pi)^{3/2}\sigma_{i}^{3}}\exp\left(-\frac{r^{2}}{2\sigma_{i}^{2}}\right)$ & -- & Fixed (MGE) \\ 
        
        Gas Disk & $\rho_{\text{gas}} \propto \exp\left(-\frac{R_m}{R} - \frac{R}{R_d}\right) \mathrm{sech}^2\left(\frac{z}{2z_d}\right)$ & -- & Fixed \\ 
        \hline
        
        \multicolumn{4}{c}{\textbf{Dark Matter Halo Models}} \\
        \hline
        NFW Halo & $\rho_{\text{NFW}}(r) = \frac{\rho_s}{(r/r_s)(1+r/r_s)^2}$ & $\log_{10}\rho_{\odot,\text{DM}}$ $[\text{GeV cm}^{-3}]$ \newline $\log_{10}r_s$ $[\text{kpc}]$ & $\mathcal{U}[-3.0, 1.0]$ \newline $\mathcal{U}[0.0, 2.3]$ \\
        
        Einasto Halo & $\rho_{\text{Ein}}(r) = \rho_s \exp\left\{-\frac{2}{\alpha_{\text{Ein}}}\left[\left(\frac{r}{r_s}\right)^{\alpha_{\text{Ein}}}-1\right]\right\}$ & $\log_{10}\rho_{\odot,\text{DM}}$ $[\text{GeV cm}^{-3}]$ \newline $\log_{10}r_s$ $[\text{kpc}]$ \newline $\alpha_{\text{Ein}}$ & $\mathcal{U}[-3.0, 1.0]$ \newline $\mathcal{U}[0.0, 2.3]$ \newline $\mathcal{U}[0, 1]$ \\
        \hline
        
        \multicolumn{4}{c}{\textbf{Modified Gravity Models}} \\
        \hline
        MOND & 
        $\nabla \cdot \left[\left(\nu\left(\frac{|\nabla\Phi_N|}{a_0}\right)-1\right)\nabla\Phi_N\right] = 4\pi G \rho_p$
        & $a_0$ \newline $[10^{-10}\text{m s}^{-2}]$ &   Fixed or $\mathcal{U}[0.3, 3.0]$ \\
        
        STVG & 
        $\begin{aligned}
            &\Phi_{\text{STVG}} = (1+\alpha_{\text{STVG}})\Phi_N + \Phi_Y \\
            &(\nabla^2 - \mu_{\text{STVG}}^2)\Phi_Y = -4\pi G \alpha_{\text{STVG}} \rho_b
        \end{aligned}$ 
        & $\alpha_{\text{STVG}}$ \newline $\mu_{\text{STVG}}$ [$\text{kpc}^{-1}$] 
        & $\mathcal{U}[7.0, 16.0]$ \newline $\mathcal{U}[0.03, 0.07]$ \\
        \hline
    \end{tabular}
\end{table}

\begin{table}
    \centering
    \setlength{\tabcolsep}{3pt} 
    \renewcommand{\arraystretch}{1.2} 
    \caption{\textbf{Summary of statistical performance and best-fit parameters for the two error scenarios.}\textbf{Observational Error Only} (Scenario 1) and \textbf{Total Error} (Scenario 2, incorporating systematic uncertainties). 
    The value of $\Delta$ (BIC and $\ln \mathcal{Z}$) are calculated relative to the best-fit model within each scenario. 
    Parameter values represent the median values with $68\%$ credible regions.  
    Gaussian priors applied in the analysis are: $\rho_{\odot,\text{thin}} = 0.045 \pm 0.003\,M_\odot \mathrm{pc}^{-3}$ and $\rho_{\odot,\text{thick}} = 0.00525 \pm 0.0006\,M_\odot \mathrm{pc}^{-3}$. Note the significant tension in modified gravity parameters in the Total Error scenario. Finally, the $\chi^2_\nu$ and BIC statistics are computed using the maximum a posteriori (MAP) parameters, which can differ from the tabulated medians for skewed distributions. Specifically, in \textbf{Scenario 2 (Total Error)}, notable deviations differ between the MAP and median values for the Einasto halo ($\alpha_{\text{Ein}} \approx 0.97$, $\log r_s \approx 0.99$, $\log \rho_{\odot,s} \approx -0.27$) and STVG ($\alpha_{\text{MOG}} \approx 10.68$, $\mu_{\text{MOG}} \approx 0.07\,\mathrm{kpc}^{-1}$). The primary results reported in this work are based on the Total Error scenario.}
    \label{tab:comprehensive_fits_final}
    \scriptsize
    \begin{tabular}{l cccc ccccccc}
        \hline
        Model & $\chi_\nu^2$ & $\Delta \mathrm{BIC}$ & $-\Delta \ln \mathcal{Z}$ & Sig. ($\sigma$) & $\rho_{\odot,\text{thin}}$ & $\rho_{\odot,\text{thick}}$ & $\log \rho_{\odot,s}$ & $\log r_s$ & $\alpha_{\text{Ein}}$ & $\alpha_{\text{MOG}}$ & $\mu_{\text{MOG}}$ \\
        & & & & & \tiny{($10^{-2}\,M_\odot \mathrm{pc}^{-3}$)} & \tiny{($10^{-3}\,M_\odot \mathrm{pc}^{-3}$)} & \tiny{(GeV cm$^{-3}$)} & \tiny{(kpc)} & & & \tiny{($10^{-2}\,\mathrm{kpc}^{-1}$)} \\
        \hline
        \multicolumn{12}{c}{\textbf{Scenario 1: Observational Error Only}} \\
        \hline
        NFW Halo      & 2.70 & 47.4 & 23.2 & 6.5 & $4.82^{+0.16}_{-0.16}$ & $5.25^{+0.58}_{-0.57}$ & $-0.40^{+0.00}_{-0.00}$ & $0.72^{+0.02}_{-0.02}$ & -- & -- & -- \\
        Einasto Halo  & \textbf{2.05} & \textbf{0.0} & \textbf{0.0} & -- & $4.59^{+0.17}_{-0.16}$ & $5.02^{+0.57}_{-0.59}$ & $-0.34^{+0.01}_{-0.01}$ & $1.04^{+0.02}_{-0.02}$ & $0.63^{+0.06}_{-0.06}$ & -- & -- \\
        \textbf{Simple} MOND & 40.54 & 2995.7 & 1464.6 & 54.0 & $7.71^{+0.13}_{-0.13}$ & $18.04^{+0.52}_{-0.52}$ & -- & -- & -- & -- & -- \\
        \textbf{Standard} MOND  & 64.01 & 4825.9 & 2400.7 & 69.2 & $13.48^{+0.14}_{-0.14}$ & $21.04^{+0.54}_{-0.54}$ & -- & -- & -- & -- & -- \\
        \textbf{RAR} MOND  & 42.04 & 3112.3 & 1524.7 & 55.1 & $8.07^{+0.13}_{-0.13}$ & $18.33^{+0.53}_{-0.53}$ & -- & -- & -- & -- & -- \\
        STVG            & 31.98 & 2273.0 & 1136.0 & 47.6 & $6.77^{+0.25}_{-0.19}$ & $13.52^{+0.55}_{-0.50}$ & -- & -- & -- & $7.01^{+0.02}_{-0.01}$ & $6.86^{+0.09}_{-0.14}$ \\
        \hline
        \multicolumn{12}{c}{\textbf{Scenario 2: Total Error}} \\
        \hline
        NFW Halo      & 1.27 & 0.7 & 0.6 & 0.9 & $4.64^{+0.20}_{-0.19}$ & $5.14^{+0.59}_{-0.58}$ & $-0.38^{+0.03}_{-0.03}$ & $0.80^{+0.13}_{-0.13}$ & -- & -- & -- \\
        Einasto Halo  & \textbf{1.22} & \textbf{0.0} & \textbf{0.0} & -- & $4.62^{+0.20}_{-0.20}$ & $5.11^{+0.60}_{-0.59}$ & $-0.31^{+0.03}_{-0.05}$ & $1.04^{+0.14}_{-0.07}$ & $0.74^{+0.18}_{-0.30}$ & -- & -- \\
        \textbf{Simple} MOND & 3.94 & 202.5 & 92.3 & 13.4 & $2.47^{+0.14}_{-0.14}$ & $5.40^{+0.56}_{-0.57}$ & -- & -- & -- & -- & -- \\
        \textbf{Standard} MOND  & 4.21 & 223.5 & 108.6 & 14.5 & $4.24^{+0.17}_{-0.17}$ & $6.02^{+0.58}_{-0.58}$ & -- & -- & -- & -- & -- \\
        \textbf{RAR} MOND  & 3.90 & 199.2 & 91.0 & 13.3 & $2.59^{+0.14}_{-0.14}$ & $5.44^{+0.58}_{-0.58}$ & -- & -- & -- & -- & -- \\
        STVG            & 1.58 & 24.0 & 10.5 & 4.2 & $4.65^{+0.19}_{-0.19}$ & $5.28^{+0.59}_{-0.58}$ & -- & -- & -- & $11.32^{+1.02}_{-0.76}$ & $6.78^{+0.17}_{-0.36}$ \\
        \hline
    \end{tabular}
\end{table}
\end{document}